\documentclass[12pt, draftclsnofoot,onecolumn, letterpaper]{IEEEtran}
\usepackage[T1]{fontenc}
\usepackage{color}
\usepackage{amsmath}
\usepackage{url}
\usepackage{array}

\usepackage{fancyhdr}
\usepackage{latexsym}
\usepackage{diagbox}
\usepackage{epsfig}
\usepackage{epstopdf}
\usepackage{amssymb}
\usepackage{amsfonts}
\usepackage{amsxtra}
\usepackage{xspace}
\usepackage{makeidx}
\usepackage{graphics}
\usepackage{theorem}
\usepackage{subfigure}
\usepackage{multirow}
\usepackage{verbatim}
\usepackage{algorithmic,algorithm}
\DeclareMathOperator*{\minimize}{minimize}

\usepackage[noadjust]{cite}

\newtheorem{lemma}{\textbf{Lemma}}

\begin{document}

\title{Channel Estimation for BIOS-Assisted Multi-User MIMO Systems: A Heterogeneous Two-timescale Strategy}

\author{Qiucen~Wu,~\IEEEmembership{Student~Member,~IEEE,}
        Tian~Lin,\IEEEmembership{}
        and~Yu~Zhu,~\IEEEmembership{Senior Member,~IEEE}

\thanks{The authors are with the Key Laboratory for Information Science of Electromagnetic Waves (MoE), School of Information Science and Technology, Fudan University, Shanghai 200433, China. (e-mail: qcwu21@m.fudan.edu.cn, lint17@fudan.edu.cn, zhuyu@fudan.edu.cn).}
\thanks{This work was supported by National Natural Science Foundation of China under Grant No. 61771147. \textit{(Corresponding author: Yu Zhu.)}}
}

\maketitle

\begin{abstract}
Bilayer intelligent omni-surface (BIOS) has recently attracted increasing attention due to its capability of independent beamforming on both reflection and refraction sides. However, its specific bilayer structure makes the channel estimation problem more challenging than the conventional intelligent reflecting surface (IRS) or intelligent omni-surface (IOS). In this paper, we investigate the channel estimation problem in the BIOS-assisted multi-user multiple-input multiple-output system. We find that in contrast to the IRS or IOS, where the forms of the cascaded channels of all user equipments (UEs) are the same, in the BIOS, those of the UEs on the reflection side are different from those on the refraction side, which is referred to as the heterogeneous channel property. By exploiting it along with the two-timescale and sparsity properties of channels and applying the manifold optimization method, we propose an efficient channel estimation scheme to reduce the training overhead in the BIOS-assisted system. Moreover, we investigate the joint optimization of base station digital beamforming and BIOS passive analog beamforming. Simulation results show that the proposed estimation scheme can significantly reduce the training overhead with competitive estimation quality, and thus keeps the performance advantage of BIOS over IRS and IOS with imperfect channel state information.

\end{abstract}

\newpage

\begin{IEEEkeywords}
Channel estimation, reconfigurable intelligent surface, bilayer-intelligent omni-surface, heterogeneous
two-timescale strategy, manifold optimization.
\end{IEEEkeywords}
\IEEEpeerreviewmaketitle

\section{Introduction}\label{sec:introduction}
Recently, reconfigurable intelligent surface (RIS) has provoked increasing attention in the evolution towards 6G communication, as it can enhance the cell coverage, provide virtual direct paths between the base station (BS) and user equipments (UEs), etc., by reconfiguring the radio propagation environment with limited hardware costs and power consumption \cite{Pan_RIS_6G_review_2021}. Typically, RIS is a passive metasurface composed of massive reconfigurable scattering elements that can control the response of impinging signals by dynamically changing the phase shift of each element so that the communication channel can be improved with specific design objectives  \cite{Smart_radio_environments_review}\cite{Zhang_RIS_review_2022}\cite{Li_RIS_review_2020}. 

\begin{figure}
	\centering
	\includegraphics[height=4.33cm,width=10cm]{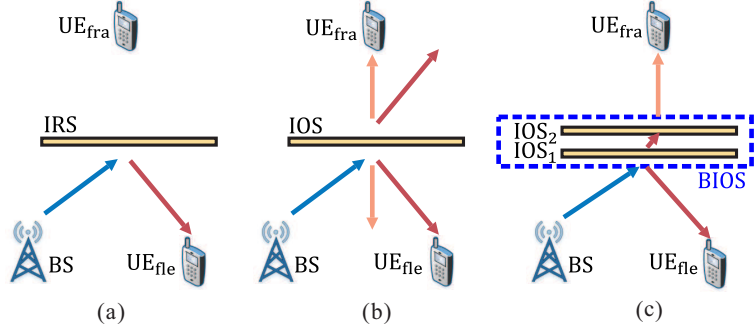}
	\caption{Examples of beamforming provided by different RIS types. (a) IRS which only serves UE$_{\mathrm{fle}}$. (b) IOS which provides symmetric beamforming on both sides. (c) BIOS which can flexibly control the beamforming on both sides.}	\label{fig:RIS modles}
\end{figure}

Most of the existing works focused on the intelligent reflecting surface (IRS), a type of RIS which can only reflect the incident signal back to the same side \cite{Renzo_IRS_review_2020}, as shown in Fig. \ref{fig:RIS modles}(a). To overcome this topological constraint of IRS, a novel RIS named as intelligent omni-surface (IOS) or simultaneous transmitting and reflecting RIS (STAR-RIS) has been recently proposed to extend the coverage of the surface to the full-space \cite{IOS_2022}\cite{STAR-RIS_2022}\footnote{For the convenience of description, we refer to both IOS and STAR-RIS as IOS in this paper.}. Specifically, as shown in Fig. \ref{fig:RIS modles}(b), the signal impinging on the IOS will be split into two parts, with one part reflected to the UEs on the same side of the incident signal (referred to as UE$_{\mathrm{fle}}$s), and the other part refracted to the UEs on the opposite side (referred to as UE$_{\mathrm{fra}}$s). Due to this unique functionality, IOS can significantly improve the performance of UEs in the whole space, and thus be employed in various communication systems such as multiple-input multiple-output (MIMO) communication \cite{STAR-RIS_MIMO_2021}, non-orthogonal multiple access  communication \cite{STAR-RIS_NOMA1_2021}\cite{STAR-RIS_NOMA2_2022}, unmanned aerial vehicle communication \cite{STAR-RIS_UAV1_2022}\cite{STAR-RIS_UAV2_2022}, etc. Meanwhile, several IOS prototypes have also been reported recently, verifying the feasibility of employing IOS in practical communication systems \cite{IOS_demo_web}\cite{STAR-RIS_prototype_2022}.

Despite of the dual functionality of reflection and refraction, the beamforming on both sides of IOS cannot be independently controlled due to the coupled phase shift for reflection and refraction signals \cite{STAR-RIS_prototype_2022}\cite{STAR_IRS_phase_couple_2021}. Specifically, in a typical case of IOS, the phase shifts for reflection and refraction signals are just the same \cite{IOS_2022}. As shown in Fig. \ref{fig:RIS modles}(b), this setup implies that the beamforming on both sides of the metasurface will be symmetric. Thus, if the UEs are randomly located in the cell, some beams are very likely not to be directed to any UE, so their power is wasted. To deal with this problem, in \cite{BIOS_2022}, we proposed a promising bilayer intelligent omni-surface (BIOS) structure with two neighbouring IOS layers (referred to as IOS$_1$ and IOS$_2$), which can flexibly control the beams on both sides of the surface, as shown in Fig. \ref{fig:RIS modles}(c). Owing to this unique capability, BIOS can provide higher data rate than IRS and IOS in multi-user systems. 

For all of the RIS types mentioned above, the acquisition of accurate channel state information (CSI) is crucial for the beamforming optimization. However, the channel estimation of RIS systems is a difficult task in practice, as all of the scattering elements are passive components without any ability of baseband signal processing. To address this issue, several RIS channel estimation schemes have been proposed in recent works. For IRS, which is the most popular type of RIS, a channel estimation scheme based on the least squares (LS) criterion for IRS-assisted single-user multiple-input single-output (MISO) systems has been proposed in \cite{LS1}, where the BS is assumed to have no prior knowledge about the channels, and thus requires high training overhead. To tackle this issue, several channel estimation approaches for reducing the overhead have been proposed by exploiting the properties of IRS channels. In particular, by utilizing the channel sparsity, several compressive sensing (CS) methods have been applied in the IRS channel estimation, e.g., orthogonal matching pursuit (OMP) \cite{OMP1}\cite{OMP2}, approximate message passing (AMP) \cite{AMP}, atomic norm minimization (ANM) \cite{ANM1}\cite{ANM2}, etc. Meanwhile, the property that all UEs share the same IRS-BS channel has also been exploited in some previous works to reduce the training overhead of channel estimation in IRS-assisted systems. For example, in \cite{yuwei}, an iterative channel estimation scheme was proposed by exploiting the fact that the sparse cascaded channels of all UEs have a common row-column-block sparsity. The authors in \cite{shared channel} utilized the correlation between UEs' reflected channels, and proposed a three-phase channel estimation scheme for an IRS-assisted multi-user MISO system. Furthermore, by reforming the received pilots as a tensor, two parallel factor based channel estimation algorithms were developed in \cite{BALS}, where the UE-IRS channels and the common IRS-BS channel are separately estimated. Another important idea to reduce the training overhead in IRS channel estimation is to utilize the channel variation property. By exploiting the fact that the coherence time periods of the IRS-BS and UE-IRS channels are different, the authors in \cite{two_timescale_channel_estimation_2021} proposed a two-timescale channel estimation scheme to separately estimate the IRS-BS and UE-IRS channels in different timescales. In addition, a three-stage estimation scheme was proposed in \cite{pan} for an IRS-assisted multi-user MIMO system, where the angles of departure (AoDs) and angles of arrival (AoAs) of channels are assumed to remain unchanged for several coherence blocks, so that only the channel gains need to be updated in these  blocks. 
%

Different from the abundant research in IRS-assisted systems, so far there has been very little research on the channel estimation in IOS-assisted systems. In \cite{STAR-RIS_channel_estimation_2021}, an LS based channel estimation scheme was proposed for the IOS-assisted system working in the time switching and energy splitting model. For the BIOS-assisted system, as there are two IOSs deployed, the channel estimation problem becomes more challenging than that in the IRS- and IOS-assisted systems. Taking the downlink transmission for example, for UE$_{\mathrm{fle}}$s, the BS signal is directly reflected by IOS$_1$ to them, while for UE$_{\mathrm{fra}}$s, the BS signal is not only refracted by the two IOSs, but also passes through the near-field channel between the two IOSs.

In this paper, we investigate the uplink channel estimation problem in the BIOS-assisted multi-user MIMO system, and propose a channel estimation scheme which can reduce the training overhead by exploiting the BIOS channel properties. To the best of our knowledge, this is the first attempt to tackle the channel estimation of the BIOS-assisted system. The main contributions of this work are summarized below:
\begin{itemize}
\item We investigate the equivalent baseband signal model of the BIOS-assisted system, and show that in contrast to the conventional IRS- and IOS-assisted systems, where the cascaded channels of all UEs have similar forms regardless of their locations, in the BIOS-assisted system the cascaded channels of UE$_{\mathrm{fle}}$s and those of UE$_{\mathrm{fra}}$s have different forms, which we refer to as the unique heterogeneous property. It is this property that makes the channel estimation in the BIOS-assisted system more difficult to deal with. 

\item By exploiting the heterogeneous and two-timescale (HTT) properties of the BIOS channels and applying the manifold optimization (MO) method, we propose the HTT-MO channel estimation scheme. Specifically, the common BIOS-BS channel among UEs is firstly estimated over a large timescale according to the uplink pilots from a UE$_{\mathrm{fra}}$ rather than a UE$_{\mathrm{fle}}$. With the estimated BIOS-BS channel, the UE-BIOS channels of all UEs are then estimated in each small timescale separately. In either the large or small timescale estimation, an MO channel estimation (MO-CE) algorithm is proposed to reduce the training overhead by exploiting the low-rank property and angle sparsity of channels.

\item With the estimated CSI, we propose a joint BS digital and BIOS analog passive beamforming optimization algorithm aiming at maximizing the downlink sum rate of the BIOS-assisted system. The original sum rate maximization problem is first converted to an equivalent weighted mean square error minimization (WMMSE) problem, and then solved by using the alternating optimization and the coordinate descent (CD) algorithm.

\item We provide various simulation results to verify the effectiveness of the proposed HTT-MO channel estimation scheme and the WMMSE-CD beamforming scheme, and show that although the channel estimation problem in the BIOS-assisted system is much more complicated than that in the conventional IRS- and IOS-assisted systems, the superiority in the sum rate performance of BIOS over IRS and IOS still holds for the situation with estimated CSI, thanks to the efficient utilization of the channel properties via the proposed HTT-MO scheme.
\end{itemize}

The rest of this paper is organized as follows. Section \ref{sec:BIOS} introduces the basic structure of BIOS. Section \ref{sec:system-model-and-channel-model} presents the system model and channel model of the BIOS-assisted multi-user MIMO system. Section \ref{sec:heterogeneous-channel-estimation-strategy} presents the HTT channel estimation strategy. Section \ref{sec:MO-EST} proposes the HTT-MO channel estimation scheme, where the detailed estimation algorithms in both the large and small timescales are discussed. Section \ref{sec:beamforming} proposes the WMMSE-CD scheme for the beamforming optimization in the BIOS-assisted system. Simulation results are provided in Section \ref{sec:Simulation Results}. Finally, Section \ref{sec:conclusion} draws the conclusions of this paper.

\emph{Notations:} In this paper, the imaginary unit is denoted by $\mathrm{j}=\sqrt{-1}$. the bold lowercase letter $\mathbf{a}$ and bold captital letter $\mathbf{A}$ represent a column vector and a matrix, respectively. $[\mathbf{a}]_i$ represents the $i$-th element of $\mathbf{a}$, and $[\mathbf{A}]_{ij}$ represents the $(i,j)$-th element of $\mathbf{A}$. $(\cdot)^T$, $(\cdot)^H$ and $(\cdot)^*$ denote the transpose, conjugate transpose and conjugate operators, respectively. $\mathrm{Tr(\cdot)}$, $\mathrm{rank(\cdot)}$ and $\mathrm{vec(\cdot)}$ denote the trace, rank and vectorization of a matrix, respectively. $\|\cdot\|_{F}$ denotes the Frobenius norm of a matrix, while $\|\cdot\|_N$ denotes the $\ell_N$-norm of a vector. $|\cdot|$ denotes the determinant (module) of a matrix (complex variable). $\mathcal{R}\{\cdot\}$ denotes the real part of a scalar. $\mathbb{E}(\cdot)$ is the expectation operator. $\circ$, $\odot$ and $\otimes$ denotes Hadamard, Khatri-Rao and Kronecker products, respectively. $\mathrm{diag}(\mathbf{a})$ denotes a diagonal matrix with the elements of $\mathbf{a}$ on its main diagonal, and $\mathrm{diag}(\mathbf{A})$ is the extraction of the diagonal of $\mathbf{A}$. $\mathrm{blkdiag}(\mathbf{A}_1,\dots,\mathbf{A}_n)$ denotes a block diagonal matrix whose diagonal components are $\mathbf{A}_1,\dots,\mathbf{A}_n$. $\mathcal{CN}(\mathbf{0}, \mathbf{K})$ denotes the circularly symmetric complex Gaussian distribution with zero mean and covariance matrix $\mathbf{K}$.
\section{Bilayer Intelligent Omni-Surface}\label{sec:BIOS}
As we mentioned in Section \ref{sec:introduction}, conventional RISs, like IRS and IOS, have several limitations in the beamforming design for multi-user systems. To overcome these limitations, we have proposed a novel RIS, BIOS, in \cite{BIOS_2022}, which consists of two neighboring layers of IOSs, as shown in Fig.~\ref{fig:RIS modles}(c). The IOS$_1$ is set in the simultaneous reflection and refraction mode, while the IOS$_2$ is set in the full penetration mode, which transmits the signal impinging on one side of it completely to the other side. Thereby, the incident signal will be split into two parts after passing through the BIOS. One is directly reflected by IOS$_1$ to UE$_{\mathrm{fle}}$s, and the other one is transmitted by IOS$_1$ to IOS$_2$, and then is refracted by IOS$_2$ to UE$_{\mathrm{fra}}$s. This unique property bestows the BIOS the degree of freedom to flexibly control the beamforming on both sides of the surface. In this paper, we assume that IOS$_1$ and IOS$_2$ are both uniform square planar arrays (UPAs) with a size of $M_\mathrm{x} \times M_\mathrm{y}$. Then, the effective coefficient matrices of BIOS for the downlink reflection and refraction signals can be expressed as
\begin{equation}\label{eqn:PHI_BIOS}
   \widehat{\mathbf{\Phi}}_\mathrm{d,fle}=\sqrt{\epsilon}{\mathbf{\Phi}}_{\mathrm{d},1},\quad \widehat{\mathbf{\Phi}}_\mathrm{d,fra}=\sqrt{1-\epsilon}\mathbf{\Phi}_{\mathrm{d},2}{\mathbf{L}^H}{\mathbf{\Phi}}_{\mathrm{d},1},
\end{equation}
where ${\mathbf{\Phi}}_{\mathrm{d},1}$ and ${\mathbf{\Phi}}_{\mathrm{d},2}\in\mathbb{C}^{M\times M}$ are the downlink diagonal coefficient matrices of IOS$_1$ and IOS$_2$, respectively, with $M=M_\mathrm{x}\times M_\mathrm{y}$. $\epsilon$ is a constant to quantify the ratio of reflection signal power to the total power of IOS$_1$, while $1-\epsilon$ is the ratio of the refraction signal power. $\mathbf{L}\in\mathbb{C}^{M\times M}$ is the near field channel matrix between IOS$_1$ and IOS$_2$, the ($m_1,m_2$) element of which can be denoted by \cite{BIOS_2022}\cite{near_channel_model} 
\begin{equation}\label{L channel model}
[{\mathbf{L}}]_{m_1m_2}=\sqrt{\frac{2a^2F(\theta^{m_1,m_2}_{\mathrm{L,t}}, \phi^{m_1,m_2}_{\mathrm{L,t}})F(\theta^{m_1,m_2}_{\mathrm{L,r}}, \phi^{m_1,m_2}_{\mathrm{L,r}})}{\pi{d^2_{m_1, m_2}}}}{\rm{exp}}\left(\frac{-\mathrm{j}2\pi{d_{m_1, m_2}}}{\lambda}\right),
\end{equation}
where $a$ is the size of IOS elements and $d_{m_1, m_2}$ is the distance between the $m_1$-th element of IOS$_1$ and the $m_2$-th element of IOS$_2$ after deployment. $F(\theta,\phi)=|\mathrm{cos}^3\theta|$ is the normalized power radiation pattern of IOS elements \cite{near_channel_model}, and $\theta^{m_1,m_2}_{\mathrm{L,t}}$($\phi^{m_1,m_2}_{\mathrm{L,t}}$), $\theta^{m_1,m_2}_{\mathrm{L,r}}$($\phi^{m_1,m_2}_{\mathrm{L,r}}$) are the elevation (azimuth) AoD and AoA of the channel between the $m_1$-th element of IOS$_1$ and the $m_2$-th element of IOS$_2$. Considering that $\mathbf{L}$ is determined only by the distance between IOS$_1$ and IOS$_2$, which remains unchanged after the deployment of BIOS, in this paper we assume that $\mathbf{L}$ is known to the BS.

\section{System Model and Channel Model}\label{sec:system-model-and-channel-model}
\subsection{System Model}
\begin{figure}
	\centering
	\includegraphics[height=5.94cm,width=9.3cm]{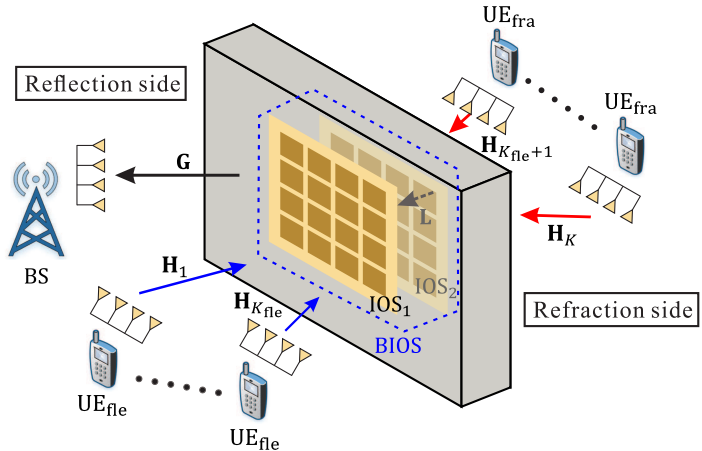}
	\caption{System model of the uplink transmission in a BIOS-assisted multi-user MIMO system.}	\label{fig:system_model}
\end{figure}
As shown in Fig. \ref{fig:system_model}, we consider the uplink channel estimation in a BIOS-assisted narrowband multi-user MIMO system operating in the time division duplex (TDD) mode, where a BS with $N_{\mathrm{BS}}$ isotropic antennas serves totally $K$ UEs equipped with $N_{\mathrm{UE}}$ isotropic antennas. As the direct links between the BS and UEs are assumed to be blocked, a BIOS is deployed to establish the virtual line-of-sight (LoS) paths for UEs. As mentioned in Section \ref{sec:BIOS}, both layers of BIOS are assumed to be UPAs consisting of $M$ reconfigurable elements. UEs are divided into two groups based on their positions relative to the BIOS, with $K_{\mathrm{fle}}$ UEs located on the reflection side, indexed by $k=1,\ldots,K_{\mathrm{fle}}$, and $K_{\mathrm{fra}} = K-K_{\mathrm{fle}}$ UEs located on the refraction side, indexed by $k={\mathrm{fle}}+1,\ldots,K$. In this work, we assume that the information of whether a UE is a UE$_{{\mathrm{fle}}}$ or a UE$_{{\mathrm{fra}}}$ is known to the BS. In order to simplify the channel estimation process and avoid the interference between pilots sent by UEs, it is assumed that the $K$ UEs send their pilots one-by-one to the BS over consecutive time. Taking the $k$-th UE for example, the equivalent baseband received signal at the BS can be represented by
\begin{equation}\label{eqn:received signal}
\begin{array}{l}
\mathbf{r}[t]=\mathbf{G}\widehat{\boldsymbol{\Phi}}_{\mathrm{\mu}(k)}[t] \mathbf{H}_{k}\mathbf{s}[t]+\mathbf{z}[t],
\end{array}
\end{equation}
where $\mathbf{s}[t]\in\mathbb{C}^{N_{\mathrm{UE}}\times 1}$ is the $t$-th pilot vector with the normalized power constraint $\big\|\mathbf{s}[t]\big\|^2=1$ and $\mathbf{z}[t]\in\mathbb{C}^{N_{\mathrm{BS}}\times 1}$ represents the additive Gaussian noise satisfying $\mathbf{z}[t]\sim\mathcal{CN}(0, \sigma^2\mathbf{I}_{N_{\mathrm{BS}}})$. $\mathrm{\mu}(k)=\mathrm{fle}$ when $k=1,\dots,K_{\mathrm{fle}}$, and $\mathrm{\mu}(k)=\mathrm{fra}$ when $k=K_{\mathrm{fle}}+1,\dots,K$. $\mathbf{\widehat{\Phi}}_{\mathrm{fle}}[t]=\sqrt{\epsilon}{\mathbf{\Phi}}_{1}[t]$ and $\mathbf{\widehat{\Phi}}_{\mathrm{fra}}[t]=\sqrt{1-\epsilon}{\mathbf{\Phi}}_{1}[t]{\mathbf{L}}\mathbf{\Phi}_{2}[t]$ are the uplink effective coefficient matrices of BIOS, with ${\mathbf{\Phi}}_{1}[t]$, ${\mathbf{\Phi}}_{2}[t]\in\mathbb{C}^{M\times M}$ denoting the uplink diagonal coefficient matrices of IOS$_1$ and IOS$_2$ for the $t$-th pilot vector. Finally, $\mathbf{G}\in\mathbb{C}^{N_{\mathrm{BS}} \times M}$ denotes the IOS$_1$-BS channel, and ${\mathbf{H}}_{k}\in\mathbb{C}^{M \times N_{\mathrm{UE}}}$ denotes the $k$-th $\mathrm{UE}_{\mathrm{fle}}$-IOS$_1$ ($k$-th $\mathrm{UE}_{\mathrm{fra}}$-IOS$_2$) channel for $k=1,\dots,K_{\mathrm{fle}}$ ($k=K_{\mathrm{fle}}+1,\dots,K$). 
 
\subsection{Cascaded Channel in the BIOS-assisted System}\label{subsec:Cascaded channel in BIOS-assisted system}

Normally, for the IRS/IOS-assisted systems, the received signal at the BS can be represented as a function of the cascaded channel of the IRS(IOS)-BS and the UE-IRS(IOS) channels for the convenience of channel estimation and beamforming design \cite{yuan}. It can be found that in the IRS/IOS-assisted systems, the forms of the cascaded channels of all UEs are the same. However, in the BIOS-assisted system, this is not the case. In particular, for UE$_{\mathrm{fle}}$s, as the $\widehat{\boldsymbol{\Phi}}_{\mathrm{fle}}[t]=\sqrt{\epsilon}{\mathbf{\Phi}_1}[t]$ is a diagonal matrix, the cascaded channel ${\mathbf{J}}_{\mathrm{fle},k}$ can be expressed as the Khatri-Rao product of $\mathbf{H}^T_{k}$ and $\mathbf{G}$, which is similar to that in the IRS/IOS-assisted MIMO system \cite{IOS_2022}\cite{ANM1}\cite{BALS}. Then, (\ref{eqn:received signal}) can be represented as
 \begin{equation}\label{eqn:received signal fle}
 \begin{array}{rll}
\mathbf{r}[t]&=&\sqrt{\epsilon}\mathbf{G}{\boldsymbol{\Phi}}_1[t] \mathbf{H}_{k}\mathbf{s}[t]+\mathbf{z}[t]\\
&=&\sqrt{\epsilon}(\mathbf{s}^T[t]\otimes\mathbf{I}_{{N}_\mathrm{BS}})(\mathbf{H}^T_{k}\odot\mathbf{G}){\boldsymbol{\phi}}_1[t] +\mathbf{z}[t],
\end{array}
\end{equation}
for $k=1,\dots,K_{\mathrm{fle}}$, where $\boldsymbol{\phi}_1[t]=\mathrm{diag}(\boldsymbol{\Phi}_1[t])$, and ${\mathbf{J}}_{\mathrm{fle},k}{=}(\mathbf{H}^T_{k}\odot\mathbf{G})\in\mathbb{C}^{N_{\mathrm{UE}}N_{\mathrm{BS}} \times M}$. (\ref{eqn:received signal fle}) follows from the facts that $\mathrm{vec}(\mathbf{ABD})=(\mathbf{D}^T\otimes\mathbf{A})\mathrm{vec}(\mathbf{B})$ and $\mathrm{vec}(\mathbf{ACD})=(\mathbf{D}^T\odot\mathbf{A})\mathbf{c}$ for any matrices $\mathbf{A}$, $\mathbf{B}$ and $\mathbf{D}$, and diagonal matrix $\mathbf{C}$ with $\mathbf{c}=\rm{diag}(\mathbf{C})$. 

For UE$_{\mathrm{fra}}$s, however, ${\boldsymbol{\Phi}}_{\mathrm{fra}}[t]=\sqrt{1-\epsilon}\mathbf{\Phi}_1[t]\mathbf{L}\mathbf{\Phi}_2[t]$ is a matrix without any zero element. Thus, (\ref{eqn:received signal}) can be rewritten as
 \begin{equation}\label{eqn:received signal fra}
 \begin{array}{rll}
\mathbf{r}[t]&=&\sqrt{1-\epsilon}\mathbf{G}\mathbf{\Phi}_1[t]\mathbf{L}\mathbf{\Phi}_2[t] \mathbf{H}_{k}\mathbf{s}[t]+\mathbf{z}[t]\\
&=&\sqrt{1-\epsilon}(\mathbf{s}^T[t]\otimes\mathbf{I}_{{N}_\mathrm{BS}})(\mathbf{H}^T_{k}\otimes\mathbf{G})\mathrm{vec}(\mathbf{\Phi}_1[t]\mathbf{L}\mathbf{\Phi}_2[t]) +\mathbf{z}[t],\\
\end{array}
\end{equation}
for $k=K_{\mathrm{fle}}+1,\dots,K$, where the cascaded channel ${\mathbf{J}}_{\mathrm{fra},k}{=}(\mathbf{H}^T_{k}\otimes\mathbf{G})\in\mathbb{C}^{N_{\mathrm{UE}}N_{\mathrm{BS}} \times M^2}$. By comparing \eqref{eqn:received signal fle} and \eqref{eqn:received signal fra}, it can be found that the forms of cascaded channels in the BS received signals are different for UE$_\mathrm{fle}$s and UE$_\mathrm{fra}$s in the BIOS-assisted system, i.e., ${\mathbf{J}}_{\mathrm{fle},k}{=}\mathbf{H}^T_{k}\odot\mathbf{G}$ for UE$_\mathrm{fle}$s and ${\mathbf{J}}_{\mathrm{fra},k}{=}\mathbf{H}^T_{k}\otimes\mathbf{G}$ and UE$_\mathrm{fra}$s. This characteristic is referred to as the heterogeneous property, which makes the channel estimation in the BIOS-assisted system more challenging. It should also be noticed that due to the heterogeneous property, the number of matrix elements in ${\mathbf{J}}_{\mathrm{fra},k}$ ($k=K_{\mathrm{fle}}+1,\dots,K$) is $M$ times of that in ${\mathbf{J}}_{\mathrm{fle},k}$ ($k=1,\dots,K_{\mathrm{fle}}$), which implies that more training overhead is required for the estimation of ${\mathbf{J}}_{\mathrm{fra},k}$. 


\subsection{Channel Model}\label{subsec: channel model}
We assume that both the BS and UEs are equipped with uniform linear array (ULA) antennas. By utilizing the Saleh-Valenzuela model \cite{yuwei} to characterize the propagation environment, for the $k$-th UE, the BIOS-BS and UE-BIOS channel matrices can be expressed as
\begin{equation}\label{eqn:channel models}
\begin{array}{cc}
      & \mathbf{G} = \sqrt{\frac{N_{\mathrm{BS}}M}{P}}\sum_{p=1}^P\sqrt{F(\theta^p_{\mathrm{t}},\varphi^p_{\mathrm{t}})}\alpha_p \mathbf{a}_{\mathrm{BS}}(\theta^p_{\mathrm{r}})\mathbf{a}^H_{\mathrm{I}}(\theta^p_{\mathrm{t}},\varphi^p_{\mathrm{t}}),\\
      &\mathbf{H}_{k} = \sqrt{\frac{N_{\mathrm{UE}}M}{Q}}\sum_{q=1}^Q\sqrt{F(\psi^q_{\mathrm{r},k},\upsilon^q_{\mathrm{r},k})}\beta_{q,k}\mathbf{a}_{\mathrm{I}}(\psi^q_{\mathrm{r},k},\upsilon^q_{\mathrm{r},k})\mathbf{a}^H_{\mathrm{UE}}(\psi^q_{\mathrm{t},k}),
\end{array}
\end{equation}
where $P$ and $Q$ denote the number of the paths of the BIOS-BS and UE-BIOS channels, respectively, which are assumed to be known by the BS. For the BIOS-BS channel, $\alpha_p$, $\theta^p_{\mathrm{r}}$ and $\theta^p_{\mathrm{t}}\ (\varphi^p_{\mathrm{t}})$ represent the complex gain, the AoA and the elevation (azimuth) AoD of the $p$-th path. For the UE-BIOS channel, similarly, $\beta_{q,k}$, $\psi^q_{\mathrm{r},k}\ (\upsilon^q_{\mathrm{r},k})$ and $\psi^q_{\mathrm{t},k}$ represent the complex gain, the elevation (azimuth) AoA and the AoD of the $q$-th path. $F(\theta^p_{\mathrm{t}},\varphi^p_{\mathrm{t}})$ \big($F(\psi^q_{\mathrm{r},k},\upsilon^q_{\mathrm{r},k})$\big) denotes the IOS normalized power radiation pattern of the $p$-th ($q$-th) path of the BIOS-BS (UE-BIOS) channel. In addition, $\mathbf{a}_{\mathrm{BS}}$, $\mathbf{a}_{\mathrm{I}}$ and $\mathbf{a}_{\mathrm{UE}}$ denote the array response vectors of the BS, BIOS and UE, respectively. Specifically, by assuming that the arrays at the BS and UEs are half-wavelength spaced ULAs and defining 
\begin{equation}\label{eqn:array response vector}
\mathbf{a}(N,x)=\frac{1}{\sqrt{N}}\left[1,e^{\mathrm{j}\pi{x}},\dots,e^{\mathrm{j}\pi(N-1){x}}\right]^T,
\end{equation}
the corresponding array response vectors of the BS and UEs can be expressed as \cite{yuwei}
\begin{equation}\label{eqn:array response vector at BS and UE}
\begin{array}{cc}
&\mathbf{a}_{\mathrm{BS}}(\theta^p_{\mathrm{r}}) = \mathbf{a}\big(N_{\mathrm{BS}},\mathrm{cos}(\theta^p_{\mathrm{r}})\big),\ \mathbf{a}_{\mathrm{UE}}(\psi^q_{\mathrm{t},k}) = \mathbf{a}\big(N_{\mathrm{UE}},\mathrm{cos}(\psi^q_{\mathrm{t},k})\big).
\end{array}
\end{equation}
For the half-wavelength spaced UPAs at BIOS, the array response vector can be given as follows
\begin{equation}\label{eqn:array response vector at BIOS}
\mathbf{a}_{\mathrm{I}}(\theta, \varphi) = \mathbf{a}\big(M_{\mathrm{x}},-\mathrm{sin}(\theta)\mathrm{sin}(\varphi)\big)\otimes\mathbf{a}\big(M_{\mathrm{y}},-\mathrm{sin}(\theta)\mathrm{cos}(\varphi)\big),
\end{equation}
where $\theta$ and $\varphi$ denote the elevation and azimuth AoD or AoA of the BIOS, respectively. 

\subsection{Sparsity of Channels in the Angle Domain}\label{subsec:Sparsity of channels in angle domain}
As mentioned in Section \ref{subsec: channel model}, there are only a few propagation paths along the UE-BIOS-BS links due to the severe path loss and blocking effects. Thus, the BIOS-BS and UE-BIOS channels can be further transformed into an angular sparse representation as follows \cite{yuwei}:
\begin{equation}\label{eqn:sparsity representation of channels}
    \begin{array}{c}
      \mathbf{G} = \mathbf{A}_{\mathrm{BS}}\mathbf{\Lambda}_{\mathbf{G}}\mathbf{A}_{\mathrm{I}}^H,\ \mathbf{H}_{k} =\mathbf{A}_{\mathrm{I}}\mathbf{\Lambda}_{\mathrm{H}, k}\mathbf{A}_{\mathrm{UE}}^H,
\end{array}
\end{equation}
where $\mathbf{A}_{\mathrm{BS}}\in\mathbb{C}^{N_{\mathrm{BS}}\times G_{\mathrm{BS}}}$, $\mathbf{A}_{\mathrm{I}}\in\mathbb{C}^{M\times G_{\mathrm{I}}}$ and $\mathbf{A}_{\mathrm{UE}}\in\mathbb{C}^{N_{\mathrm{UE}}\times G_{\mathrm{UE}}}$ are the overcomplete angular domain dictionaries with the angle resolutions of $G_{\mathrm{BS}}$, $G_{\mathrm{I}}$ and $G_{\mathrm{UE}}$, respectively. Each column of the dictionaries corresponds to one specific AoA/AoD at the BS, BIOS or the $k$-th UE. $\mathbf{\Lambda}_{\mathbf{G}}\in\mathbb{C}^{G_{\mathrm{BS}}\times G_{\mathrm{I}}}$ and $\mathbf{\Lambda}_{\mathrm{H}, k}\in\mathbb{C}^{G_{\mathrm{I}}\times G_{\mathrm{UE}}}$ are the angular domain sparse matrices of $\mathbf{G}$ and $\mathbf{H}_{k}$, which consist of $P$ and $Q$ non-zero elements respectively corresponding to the channel path gains. By selecting the codewords from the uniform grid, $\mathbf{A}_{\mathrm{BS}}$ and $\mathbf{A}_{\mathrm{UE}}$ can be expressed as
\begin{equation}\label{eqn:dictionary1}
\begin{array}{l}
       \mathbf{A}_{\mathrm{BS}} = \left[\mathbf{a}\big(N_{\mathrm{BS}},x^1_\mathrm{BS}\big),\dots,\mathbf{a}\big(N_{\mathrm{BS}},x^{G_{\mathrm{BS}}}_\mathrm{BS}\big)\right], \  \mathbf{A}_{\mathrm{UE}} = \left[\mathbf{a}\big(N_{\mathrm{UE}},x^1_\mathrm{UE}\big),\dots,\mathbf{a}\big(N_{\mathrm{UE}},x^{G_{\mathrm{UE}}}_\mathrm{UE}\big)\right],
\end{array}
\end{equation}
where $x^i_\mathrm{BS} = -1 + (i-1)\frac{2}{G_\mathrm{BS}}$ and $x^i_\mathrm{UE} = -1 + (i-1)\frac{2}{G_\mathrm{UE}}$. By setting $G_{\mathrm{x}}$ and $G_{\mathrm{y}}$ to be the angular resolutions of the BIOS along the x-axis and y-axis, $\mathbf{A}_{\mathrm{I}}$ can be exhibited in a similar way as $\mathbf{A}_{\mathrm{I}} = \mathbf{A}_{\mathrm{x}}\otimes\mathbf{A}_{\mathrm{y}}$, where 
\begin{equation}\label{eqn:AyAz}
\mathbf{A}_{\mathrm{x}} = \left[\mathbf{a}\big(M_{\mathrm{x}},x^1_\mathrm{x}\big),\dots,\mathbf{a}\big(M_{\mathrm{x}},x^{G_{\mathrm{x}}}_\mathrm{x}\big)\right], \mathbf{A}_{\mathrm{y}} = \left[\mathbf{a}\big(M_{\mathrm{y}},x^1_\mathrm{y}\big),\dots,\mathbf{a}\big(M_{\mathrm{y}},x^{G_{\mathrm{y}}}_\mathrm{y}\big)\right]
\end{equation}
with $x^i_\mathrm{x} = -1 + (i-1)\frac{2}{G_\mathrm{x}}$, $x^i_\mathrm{y} = -1 + (i-1)\frac{2}{G_\mathrm{y}}$ and $G_{\mathrm{x}}G_{\mathrm{y}} = G_{\mathrm{I}}$.

\section{Heterogeneous Two-timescale Channel Estimation Strategy}\label{sec:heterogeneous-channel-estimation-strategy}
As we mentioned in Section \ref{sec:system-model-and-channel-model}, the number of coefficients in the cascaded channels is huge, especially for $\mathbf{J}_{\mathrm{fra},k}$, which leads to prohibitive training overhead if we directly estimate the cascaded channels of all UEs. In this section, we propose an HTT channel estimation strategy to reduce the requested training overhead by exploiting the channel properties in the BIOS-assisted system.

\subsection{The Two-timescale Property} \label{subsection:The shared and two-timescale properties of channels in BIOS-assisted system}

It can be observed that in a BIOS-assisted system, the time variation of the BIOS-BS channel and that of the UE-BIOS channel are in different scales. In particular, on one hand, as the BS and BIOS are rarely moved after deployment, the BIOS-BS channel $\mathbf{G}$, which is shared by all UEs, can be regarded as unchanged for a long period of time. On the other hand, the UE-BIOS channel $\mathbf{H}_k$ varies in a much smaller timescale, since the movement of UEs frequently changes the propagation geometry between UEs and BIOS \cite{two_timescale_channel_estimation_2021}. This two-timescale property allows the BS to only estimate $\mathbf{G}$ once in the large timescale, and with the estimated $\mathbf{G}$, then $\mathbf{H}_k$ can be further estimated in the small timescale. Thus, the training overhead can be significantly reduced. 

However, the main difficulty is how to estimate $\mathbf{G}$, as the scattering elements in the BIOS-assisted system (and also the conventional RIS systems) are passive components without any capability of baseband processing. To deal with this difficulty, the authors in \cite{two_timescale_channel_estimation_2021} proposed a dual-link pilot transmission scheme for the IRS-assisted system, where the full-duplex BS estimates the BS-IRS channel by firstly sending pilots to the IRS through the downlink channel, and then receiving the reflected signals from the IRS via the uplink channel. Although this scheme provides a way to estimate the BS-IRS channel matrix, the requirement of full duplex mode places high demands on the BS hardware. In this paper, we show that it is possible to estimate the BIOS-BS channel $\mathbf{G}$ without taking the full-duplex assumption, but based on the uplink pilots sent by a UE in the large timescale, and the estimated $\mathbf{G}$ can be used for the estimation of the UE-BIOS channels of all UEs. However, to make it workable, we need to deal with the challenge of the heterogeneous property of the BIOS channels. 



\subsection{The Challenge of the Heterogeneous Property}
\label{subsection:Recovery of G and H from the cascaded channel}
As we have pointed out in Section \ref{subsec:Cascaded channel in BIOS-assisted system} that, from (\ref{eqn:received signal fle}) and (\ref{eqn:received signal fra}), the signals received by the BS from the UE$_{{\mathrm{fle}}}$s and the UE$_{{\mathrm{fra}}}$s have different expressions of the cascaded channels, where the former is a function of ${\mathbf{J}}_{\mathrm{fle},k}{=}\mathbf{H}^T_{k}\odot\mathbf{G}$ and the latter is a function of ${\mathbf{J}}_{\mathrm{fra},k}{=}\mathbf{H}^T_{k}\otimes\mathbf{G}$. This channel heterogeneous property makes the channel estimation in the BIOS-assisted system more challenging, especially when taking the two-timescale property to reduce the training overhead. Taking a UE$_{{\mathrm{fle}}}$ for example and even assuming that there is no noise effect in the channel estimation, the best channel estimates the BS can obtain, denoted by $\widetilde{\mathbf{G}}$ and $\widetilde{\mathbf{H}}_{k}$, are not exactly $\mathbf{G}$ and $\mathbf{H}_k$, but belong to a set satisfying the condition of ${\mathbf{J}}_{\mathrm{fle},k}{=}\mathbf{H}^T_{k}\odot\mathbf{G}=\widetilde{\mathbf{H}}^T_{k}\odot\widetilde{\mathbf{G}}$. Similarly, for a UE$_{{\mathrm{fra}}}$, the best channel estimates belong to a set satisfying ${\mathbf{J}}_{\mathrm{fra},k}{=}\mathbf{H}^T_{k}\otimes\mathbf{G}=\widetilde{\mathbf{H}}^T_{k}\otimes\widetilde{\mathbf{G}}$. Obviously due to the heterogeneous property, these two sets are different. However, in the two-timescale strategy, $\mathbf{G}$ needs to be estimated first and used for the estimation of $\mathbf{H}^T_{k}$ of all UEs. Thus, a workable channel estimation scheme must be the one that can make $\widetilde{\mathbf{G}}$ belong to both sets or at least their subsets. The following lemmas provide detailed mathematical proofs.

\begin{lemma}\label{lem:freedom of G and H}
If there is no noise effect in the channel estimation, the channel estimates  $\widetilde{\mathbf{G}}$ and $\widetilde{\mathbf{H}}_{k}$ should belong to one of the following two sets depending on whether the UE is a UE$_{{\mathrm{fle}}}$ or a UE$_{{\mathrm{fra}}}$
\begin{equation}\label{eqn:set of G and H for fle}
    (\widetilde{\mathbf{H}}_{k}^T, \widetilde{\mathbf{G}}) \in\  \mathcal{S}_{\mathrm{fle}, k}=\left\{(\mathbf{A}, \mathbf{B})\ |\ \mathbf{A}\odot \mathbf{B}=\mathbf{H}_{k}^T\odot \mathbf{G}\right\}, \ k=1,\dots,K_{\mathrm{fle}},\ \ \ \ \ \ 
\end{equation}
\begin{equation}\label{eqn:set of G and H for fra}
    (\widetilde{\mathbf{H}}_{k}^T, \widetilde{\mathbf{G}}) \in\  \mathcal{S}_{\mathrm{fra}, k}=\left\{(\mathbf{A}, \mathbf{B})\ |\ \mathbf{A}\otimes \mathbf{B}=\mathbf{H}_{k}^T\otimes \mathbf{G}\right\},\ k=K_{\mathrm{fle}}+1,\dots,K,
\end{equation}
which can be further proved to be equal to the following forms respectively
\begin{equation}\label{eqn:set of G and H fle2}
\mathcal{S}_{\mathrm{fle},k}=\Big\{(\mathbf{A}, \mathbf{B})\Big|\ \exists\  a_1,\dots,a_M\neq0, \mathbf{A}= [a_1\mathbf{h}_1, \dots, a_M\mathbf{h}_M], \mathbf{B}=\Big[{\frac{1}{a_1}}\mathbf{g}_1, \dots, \frac{1}{a_M}\mathbf{g}_M\Big]\Big\}, 
\end{equation}
\begin{equation}\label{eqn:set of G and H fra2}
\mathcal{S}_{\mathrm{fra},k}=\Big\{(\mathbf{A}, \mathbf{B})\Big|\ \exists\  a \neq0,\ \mathbf{A}=a\mathbf{H}_{ k}^T,\ \mathbf{B}=\frac{1}{a} \mathbf{G}\Big\},
\end{equation}
where $\mathbf{H}_{ k}^T=[\mathbf{h}_1, \dots,\mathbf{h}_M]$, $\mathbf{G}=[\mathbf{g}_1, \dots,\mathbf{g}_M]$.
\end{lemma}
\emph{Proof}: (\ref{eqn:set of G and H for fle}) and (\ref{eqn:set of G and H for fra}) can be directly proved from the definitions of the cascaded channels in (\ref{eqn:received signal fle}) and (\ref{eqn:received signal fra}), and the proofs of (\ref{eqn:set of G and H fle2}) and (\ref{eqn:set of G and H fra2}) can be found in Appendix \ref{app:equivalence between sets}.$\hfill\blacksquare$

As can be observed in the Lemma \ref{lem:freedom of G and H}, for UE$_{\mathrm{fle}}$s, each column of the channel estimate $\widetilde{\mathbf{G}}$ can differ from the corresponding column of the true channel matrix ${\mathbf{G}}$ with a distinct non-zero coefficient. However, for UE$_{\mathrm{fra}}$s, these non-zero coefficients should be identical for all columns, which means the feasible set of $\widetilde{\mathbf{G}}$ for UE$_{\mathrm{fra}}$s is a subset of that for UE$_{\mathrm{fle}}$s. Thus, it can be shown in the following lemma that, to implement the two-timescale strategy in the BIOS-assisted system, if a UE is selected to send pilots for the BS to estimate ${\mathbf{G}}$ in the large timescale and then use the result to estimate ${\mathbf{H}}^T_k$ of all UEs in the small timescale, such a UE should be a UE$_\mathrm{fra}$ rather than a UE$_{\mathrm{fle}}$ due to the heterogeneous property.

\begin{lemma}\label{lem:fra for fle}
If a UE$_{\mathrm{fra}}$ is chosen to send pilots to the BS for the estimation of $\mathbf{G}$ in the large timescale and  $\widehat{\mathbf{G}}$ is defined as the estimation result with no noise effect, then for UE$_{\mathrm{fra}}$s, $(\widehat{\mathbf{H}}^T_k, \widehat{\mathbf{G}})$ belongs to $\mathcal{S}_{\mathrm{fra},k}$, while for UE$_{\mathrm{fle}}$s,  $(\widehat{\mathbf{H}}^T_k, \widehat{\mathbf{G}})$ belongs to a subset of $\mathcal{S}_{\mathrm{fle},k}$ with $a_1=\dots=a_M$, where  $\widehat{\mathbf{H}}_k$ is the estimation result in the small timescale with no noise effect.

However, if a UE$_{\mathrm{fle}}$ is chosen to send pilots in the large timescale, then for UE$_{\mathrm{fra}}$s, the resulting $(\widehat{\mathbf{H}}^T_k, \widehat{\mathbf{G}})$ does not belong to $\mathcal{S}_{\mathrm{fra},k}$, which means that the accurate CSI for UE$_{\mathrm{fra}}$s cannot be achieved even when there is no noise effect.





\end{lemma}
\emph{Proof}: To prove the first part of this lemma, if a UE$_{\mathrm{fra}}$ is chosen to send pilots to the BS for the estimation of the true channel ${\mathbf{G}}$ in the large timescale, then $\widehat{\mathbf{G}}$ can be expressed as $\widehat{\mathbf{G}}=\frac{1}{a}{\mathbf{G}}$ according to (\ref{eqn:set of G and H fra2}). With $\widehat{\mathbf{G}}$, the BS can estimate ${\mathbf{H}}_k$ UE-by-UE in the small timescale, based on its received pilots sent from UEs, and the estimation result $\widehat{\mathbf{H}}_k$ for all UEs should satisfy  $\widehat{\mathbf{H}}_k=a{\mathbf{H}_k}$ according to (\ref{eqn:set of G and H fle2}) and (\ref{eqn:set of G and H fra2}). Then, it can be seen that for UE$_{\mathrm{fra}}$s,  $(\widehat{\mathbf{H}}^T_k, \widehat{\mathbf{G}})$ belongs to $\mathcal{S}_{\mathrm{fra},k}$, while for UE$_{\mathrm{fle}}$s,  $(\widehat{\mathbf{H}}^T_k, \widehat{\mathbf{G}})$ is in a subset of $\mathcal{S}_{\mathrm{fle},k}$ with $a_1=\dots=a_M$.

To prove the second part, if a UE$_{\mathrm{fle}}$ is chosen to send pilots for the estimation of ${\mathbf{G}}$ in the large timescale, the resulting $\widehat{\mathbf{G}}$ can be represented as $\widehat{\mathbf{G}}=\Big[{\frac{1}{a_1}}\mathbf{g}_1, \dots, \frac{1}{a_M}\mathbf{g}_M\Big]$ according to (\ref{eqn:set of G and H fle2}), where $a_i$ is a distinct coefficient for each column. However, based on this $\widehat{\mathbf{G}}$, in the small timescale, for UE$_{\mathrm{fra}}$s, the BS cannot find a channel estimate $\widehat{\mathbf{H}}_{k}$ so that $(\widehat{\mathbf{H}}_{k}^T, \widehat{\mathbf{G}})\in\mathcal{S}_{\mathrm{fra}, k}$, i.e., $\widehat{\mathbf{H}}_{k}^T\otimes \widehat{\mathbf{G}}\neq\mathbf{H}_{k}^T\otimes \mathbf{G}$, since $\widehat{\mathbf{H}}_{k}^T=a\widehat{\mathbf{H}}_{k}$ according to (\ref{eqn:set of G and H fra2}) but $\widehat{\mathbf{G}}=\Big[{\frac{1}{a_1}}\mathbf{g}_1, \dots, \frac{1}{a_M}\mathbf{g}_M\Big]$. $\hfill\blacksquare$
\subsection{HTT Channel Estimation Strategy}\label{subsec:heterogeneous-channel-estimation-strategy}
\begin{figure}
	\centering
	\includegraphics[height=4.52cm,width=11.4cm]{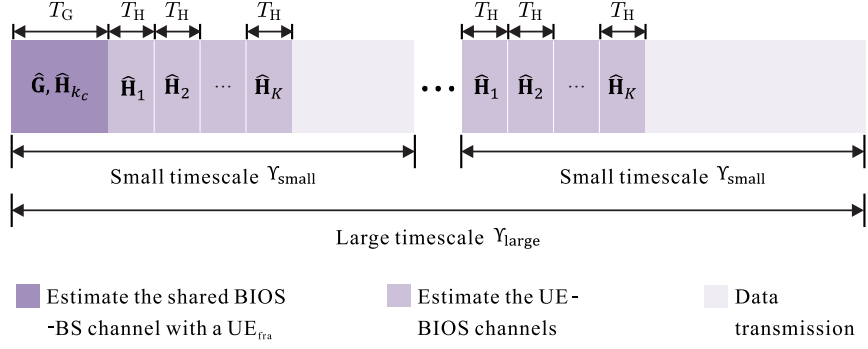}
	\caption{The proposed HTT channel estimation strategy.}	\label{two step strategy}
\end{figure}

With the two-timescale and heterogeneous properties of the BIOS channels discussed above, we come up to the HTT channel estimation strategy, which is also depicted in Fig. \ref{two step strategy}. At first, the BS estimates the BIOS-BS channel over a large timescale based on the $T_\mathrm{G}$ uplink pilots sent by a selected UE$_{\mathrm{fra}}$, e.g., the $k_c$-th UE. Then, in the small timescale, the BS estimates the UE-BIOS channel for each UE based on the received $T_\mathrm{H}$ uplink pilots and its estimated channel matrix $\widehat{\mathbf{G}}$ in the large timescale. It can be found that in the large timescale, there are a total of $MN_{\mathrm{UE}}+MN_{\mathrm{BS}}$ coefficients need to be estimated for ${\mathbf{G}}$ and ${\mathbf{H}}_{k_c}$, while in the small timescale, there are a total of $KMN_{\mathrm{UE}}$ coefficients in ${\mathbf{H}}_{k}$ for all UEs. In comparison, in the conventional cascaded channel estimation strategy, the BS needs to estimate the cascaded channel for each UE, and thus the total number of coefficients need to be estimated is $K_{\mathrm{fle}}MN_{\mathrm{UE}}N_{\mathrm{BS}}+K_{\mathrm{fra}}M^2N_{\mathrm{UE}}N_{\mathrm{BS}}$. Comparing these two numbers, it can be seen that the HTT channel estimation strategy can significantly reduce the number of coefficients to be estimated and in turn the training overhead.

\section{HTT-MO Channel Estimation Scheme}\label{sec:MO-EST}
In this section, based on the proposed HTT channel estimation strategy, we exploit the channel sparsity to further reduce the pilot overhead, and design the HTT-MO scheme for the channel estimation in the BIOS-assisted system. 

\subsection{Channel Sparsity}\label{subsec:Sparsity of channels}
Although the proposed HTT strategy can significantly reduce the training overhead, the channel estimation is still challenging due to the large number of antennas at both the BS and UEs, and the large number of scattering elements at the BIOS as well, which result in much training overhead. Fortunately, it can be further reduced by exploiting the channel sparsity in the angle domain shown in (\ref{eqn:sparsity representation of channels}). The following two lemmas illustrate the channel sparsity of the BIOS-assisted system in detail.
\begin{lemma}\label{lem:low rank of channel}
If $\mathrm{min}\{N_{\mathrm{BS}},M\}\ge P$ and $\mathrm{min}\{N_{\mathrm{UE}},M\}\ge Q$, where $P$ and $Q$ denote the number of paths of the BIOS-BS and UE-BIOS channels, respectively, as defined in \eqref{eqn:channel models}, we have
\begin{equation}\label{[equ:rank of channel}
    \mathrm{rank}(\mathbf{G}) = P,\ \mathrm{rank}(\mathbf{H}_k) = Q.
\end{equation}
\end{lemma}
\emph{Proof}: We take the proof of $\mathrm{rank}(\mathbf{G}) = P$ as an example. The proof of $\mathrm{rank}(\mathbf{H}_k) = Q$ can be completed in a similar way. According to (\ref{eqn:channel models}), the BIOS-BS channel $\mathbf{G}$ can be rewritten as 
\begin{equation}\label{eqn:rewritten of G}
    \mathbf{G} = \overline{\mathbf{A}}_\mathrm{BS}\overline{\mathbf{\Lambda}}_{\mathbf{G}}\overline{\mathbf{A}}_\mathrm{I}^H,
\end{equation}
where $\overline{\mathbf{A}}_\mathrm{BS}=[\mathbf{a}_{\mathrm{BS}}({{\theta}^1_{\mathrm{r}}}),\dots,\mathbf{a}_{\mathrm{BS}}({\theta}^P_{\mathrm{r}})]$, $\overline{\mathbf{\Lambda}}_\mathbf{G}=\sqrt{\frac{N_{\mathrm{BS}}M}{P}} \mathrm{diag}(\alpha_1,\dots,\alpha_P)$ and $\overline{\mathbf{A}}_\mathrm{I}=\overline{\mathbf{A}}_{\mathrm{x}}\odot \overline{\mathbf{A}}_{\mathrm{y}}$, with $\overline{\mathbf{A}}_{\mathrm{x}}=[\mathbf{a}\big(M_{\mathrm{x}},-\mathrm{sin}(\theta^1_{\mathrm{t}})\mathrm{sin}(\varphi^1_{\mathrm{t}})\big),\dots,\mathbf{a}\big(M_{\mathrm{x}},-\mathrm{sin}(\theta^P_{\mathrm{t}})\mathrm{sin}(\varphi^P_{\mathrm{t}})\big)]$ and $\overline{\mathbf{A}}_{\mathrm{y}}=[\mathbf{a}\big(M_{\mathrm{y}},-\mathrm{sin}(\theta^1_{\mathrm{t}})$\\$\times\mathrm{cos}(\varphi^1_{\mathrm{t}})\big),\dots,\mathbf{a}\big(M_{\mathrm{y}},-\mathrm{sin}(\theta^P_{\mathrm{t}})\mathrm{cos}(\varphi^P_{\mathrm{t}})\big)]$. As all of the column vectors of $\overline{\mathbf{A}}_\mathrm{BS}$ are linearly independent, we have $\mathrm{rank}(\overline{\mathbf{A}}_\mathrm{BS}) = P$. Similarly, we can also obtain that $\mathrm{rank}(\overline{\mathbf{A}}_\mathrm{x}) = P$ and $\mathrm{rank}(\overline{\mathbf{A}}_\mathrm{y}) = P$. According to the
properties of Kronecker product and Khatri-Rao matrix product, we can transform $\overline{\mathbf{A}}_\mathrm{I}$ into the following form
\begin{equation}\label{eqn:rewritten of A_BIOS}
    \overline{\mathbf{A}}_\mathrm{I} = \overline{\mathbf{A}}_{\mathrm{x}}\odot \overline{\mathbf{A}}_{\mathrm{y}} = \left(\overline{\mathbf{A}}_{\mathrm{x}}\otimes \overline{\mathbf{A}}_{\mathrm{y}}\right)\left(\mathbf{I}_{P}\odot\mathbf{I}_{P}\right),
\end{equation}
where $\mathrm{rank}(\overline{\mathbf{A}}_{\mathrm{x}}\otimes \overline{\mathbf{A}}_{\mathrm{y}})=\mathrm{rank}(\overline{\mathbf{A}}_{\mathrm{x}})\mathrm{rank}(\overline{\mathbf{A}}_{\mathrm{y}})=P^2$ and $\mathrm{rank}(\mathbf{I}_{P}\odot\mathbf{I}_{P}) = P$. Then, according to the rank properties of matrix \cite{matrix property}, for arbitrary $\mathbf{A}\in\mathbb{C}^{m\times k}$ and $\mathbf{B}\in\mathbb{C}^{k\times n}$, we have
\begin{equation}\label{eqn:ranke properties}
\begin{array}{l}
    \mathrm{rank}\left({\mathbf{A}}{\mathbf{B}}\right) \le\mathrm{min}\{\mathrm{rank}\left({\mathbf{A}}\right),\mathrm{rank}\left({\mathbf{B}}\right)\},\ 
    \mathrm{rank}\left({\mathbf{A}}{\mathbf{B}}\right) \ge\mathrm{rank}\left({\mathbf{A}}\right) + \mathrm{rank}\left({\mathbf{B}}\right) - k.
\end{array}
\end{equation}
By substituting (\ref{eqn:rewritten of A_BIOS}) into (\ref{eqn:ranke properties}), we have $\mathrm{rank}(\overline{\mathbf{A}}_\mathrm{I})=P$. Finally, as the rank of $\overline{\mathbf{\Lambda}}_\mathbf{G}$ is also $P$, we can further substitute (\ref{eqn:rewritten of G}) into (\ref{eqn:ranke properties}) and obtain that $\mathrm{rank}({\mathbf{G}})=P$.$\hfill\blacksquare$

\begin{lemma}\label{lem: sparsity of angle}
If $G_{\mathrm{UE}}=N_{\mathrm{UE}}$, $G_{\mathrm{BS}}=N_{\mathrm{BS}}$ and $G_{\mathrm{I}}=M$, we have $\left\|\boldsymbol{\lambda}_{\mathbf{G}}\right\|_{0}=P$, $\left\|\boldsymbol{\lambda}_{\mathbf{H}_k}\right\|_{0}=Q$,
with $\boldsymbol{\lambda}_{\mathbf{G}}=\mathrm{vec}(\mathbf{A}_{\mathrm{BS}}^H\mathbf{G}\mathbf{A}_{\mathrm{I}})$ and $\boldsymbol{\lambda}_{\mathbf{H}_k}=\mathrm{vec}(\mathbf{A}_{\mathrm{I}}^H\mathbf{H}_k\mathbf{A}_{\mathrm{UE}})$.
\end{lemma}
\emph{Proof}: We take the proof of $\left\|\boldsymbol{\lambda}_{\mathbf{G}}\right\|_{0}=P$ as an example. According to Section \ref{subsec:Sparsity of channels in angle domain}, $\mathbf{A}_{\mathrm{BS}}$ and $\mathbf{A}_{\mathrm{I}}$ are both unitary matrices as $G_{\mathrm{BS}}=N_{\mathrm{BS}}$ and $G_{\mathrm{I}}=M$. Therefore, $\boldsymbol{\lambda}_{\mathbf{G}}$ can be rewritten in the form  $\boldsymbol{\lambda}_{\mathbf{G}}=\mathrm{vec}(\mathbf{A}_{\mathrm{BS}}^H\mathbf{G}\mathbf{A}_{\mathrm{I}})=\mathrm{vec}(\mathbf{A}_{\mathrm{BS}}^H\mathbf{A}_{\mathrm{BS}}\mathbf{\Lambda}_{\mathbf{G}}\mathbf{A}_{\mathrm{I}}^H\mathbf{A}_{\mathrm{I}})=\mathrm{vec}(\mathbf{\Lambda}_{\mathbf{G}})$. As $\mathbf{\Lambda}_{\mathbf{G}}$ only consists of ${P}$ non-zero elements, we have $\left\|\boldsymbol{\lambda}_{\mathbf{G}}\right\|_{0}=\left\|\mathrm{vec}(\mathbf{\Lambda}_{\mathbf{G}})\right\|_{0}=P$.$\hfill\blacksquare$

Lemma \ref{lem:low rank of channel} and Lemma \ref{lem: sparsity of angle} reveal the channel low-rank and angle sparse properties, respectively. Although these properties correspond to the true channel matrices $\mathbf{G}$ and $\mathbf{H}_k$, it can be also proved that for any UE, Lemma \ref{lem:low rank of channel} and Lemma \ref{lem: sparsity of angle} still hold for any estimated $\widehat{\mathbf{G}}$ and $\widehat{\mathbf{H}}_{k}$, as according to Lemma 2, there is only a scalar difference between $\widehat{\mathbf{G}}$($\widehat{\mathbf{H}}_{k}$) and ${\mathbf{G}}$(${\mathbf{H}}_{k}$).

\subsection{Estimation of $\mathbf{G}$ in the Large Timescale} \label{subsec:Estimation of G in large timescale by MO algorithm}
For the two-timescale strategy and according to Lemma \ref{lem:fra for fle}, the first step at the BS is to estimate $\mathbf{G}$ based on its received pilots sent from a UE$_{\mathrm{fra}}$, e.g., the $k_c$-th UE for $K_{\mathrm{fle}}+1\leq k_c\leq K$, in the large timescale. Based on \eqref{eqn:received signal fra}, the channel estimation objective can be established similar to that of the LS problem. By further utilizing the channel sparsity properties exhibited in Lemmas \ref{lem:low rank of channel} and \ref{lem: sparsity of angle}, the estimation problem of $\mathbf{G}$ along with $\mathbf{H}_{k_c}$ can be formulated as
\begin{equation}\label{prb:LS estimation}
\begin{array}{cl}
\underset{\widehat{\mathbf{G}}, \widehat{\mathbf{H}}_{k_c}}{\operatorname{minimize}} & \sum_{t=1}^{T_\mathbf{G}}\Big\|\mathbf{r}[t]-\sqrt{1-\epsilon}\widehat{\mathbf{G}}\mathbf{\Phi}_1[t]\mathbf{L}\mathbf{\Phi}_2[t] \widehat{\mathbf{H}}_{ k_c}\mathbf{s}[t]\Big\|^{2} \\
\text { subject to } & \operatorname{rank}(\widehat{\mathbf{G}})=P, \quad \operatorname{rank}(\widehat{\mathbf{H}}_{k_c})=Q, \\
& \big\|\boldsymbol{\lambda}_{\widehat{\mathbf{G}}}\big\|_{0}=P, \quad\big\|\boldsymbol{\lambda}_{\widehat{\mathbf{H}}_{k_c}}\big\|_{0}=Q,
\end{array}
\end{equation}
where $\boldsymbol{\lambda}_{\widehat{{\mathbf{G}}}}=\mathrm{vec}(\mathbf{A}_{\mathrm{BS}}^H\widehat{\mathbf{G}}\mathbf{A}_{\mathrm{I}})$, $\boldsymbol{\lambda}_{\widehat{\mathbf{H}}_{k_c}}=\mathrm{vec}(\mathbf{A}_{\mathrm{I}}^H\widehat{\mathbf{H}}_{k_c}\mathbf{A}_{\mathrm{UE}})$, and $T_\mathbf{G}$ is the number of pilot vectors for the estimation of $\mathbf{G}$ in one large timescale. It can be seen that (\ref{prb:LS estimation}) is difficult to solve due to the multiple coupled variables and the highly non-convex constraints, and thus it may not be possible to achieve a globally optimal solution. However, with some specific processing, we can rewrite the original problem and obtain a locally optimal solution of $\widehat{\mathbf{G}}$. First to deal with the $\ell_{0}$-norm constraints in \eqref{prb:LS estimation}, we use the $\ell_{1}$-norm regularization to relax them and rewrite \eqref{prb:LS estimation} as 
\begin{equation}\label{prb:LS estimation l1 norm}
\begin{array}{cl}
\underset{\widehat{\mathbf{G}}, \widehat{\mathbf{H}}_{k_c}}{\operatorname{minimize}} & \sum_{t=1}^{T_\mathbf{G}}\big\|\mathbf{r}[t]-\sqrt{1-\epsilon}\widehat{\mathbf{G}}\mathbf{\Phi}_1[t]\mathbf{L}\mathbf{\Phi}_2[t] \widehat{\mathbf{H}}_{k_c}\mathbf{s}[t]\big\|^{2} \\
& + \upsilon_{\widehat{\mathbf{G}}}\big\|\boldsymbol{\lambda}_{\widehat{\mathbf{G}}}\big\|_{1}+ \upsilon_{\widehat{\mathbf{H}}_{k_c}}\big\|\boldsymbol{\lambda}_{\widehat{\mathbf{H}}_{k_c}}\big\|_{1}\\
\text { subject to } & \operatorname{rank}(\widehat{\mathbf{G}})=P, \quad \operatorname{rank}(\widehat{\mathbf{H}}_{k_c})=Q, 
\end{array}
\end{equation}
where $\upsilon_{\widehat{\mathbf{G}}}$ and $\upsilon_{\widehat{\mathbf{H}}_{k_c}}$ are the tuning parameters which control the contributions of the $\ell_{1}$-norm. To further deal with the multiple coupled variables $\widehat{\mathbf{G}}$ and $\widehat{\mathbf{H}}_{k_c}$, we apply the alternating minimization method and decompose (\ref{prb:LS estimation l1 norm}) into the following two subproblems
\begin{equation}\label{prb:LS estimation l1 norm G}
\begin{array}{ll}
\underset{\widehat{\mathbf{G}}}{\operatorname{minimize}} & f_{\widehat{\mathbf{G}}}\ =\sum_{t=1}^{T_\mathbf{G}}\big\|\mathbf{r}[t]-\sqrt{1-\epsilon}\widehat{\mathbf{G}}\mathbf{\Phi}_1[t]\mathbf{L}\mathbf{\Phi}_2[t] \widehat{\mathbf{H}}_{k_c}\mathbf{s}[t]\big\|^{2}  + \upsilon_{\widehat{\mathbf{G}}}\big\|\boldsymbol{\lambda}_{\widehat{\mathbf{G}}}\big\|_{1}\ \ \ \ \ \\
\text { subject to } & \operatorname{rank}(\widehat{\mathbf{G}})=P, 
\end{array}
\end{equation}

\begin{equation}\label{prb:LS estimation l1 norm H}
\begin{array}{ll}
\underset{\widehat{\mathbf{H}}_{k_c}}{\operatorname{minimize}} & f_{\widehat{\mathbf{H}}_{k_c}}=\sum_{t=1}^{T_\mathbf{G}}\big\|\mathbf{r}[t]-\sqrt{1-\epsilon}\widehat{\mathbf{G}}\mathbf{\Phi}_1[t]\mathbf{L}\mathbf{\Phi}_2[t] \widehat{\mathbf{H}}_{ k_c}\mathbf{s}[t]\big\|^{2}  + \upsilon_{\widehat{\mathbf{H}}_{k_c}}\big\|\boldsymbol{\lambda}_{\widehat{\mathbf{H}}_{k_c}}\big\|_{1}\\
\text { subject to } &  \operatorname{rank}(\widehat{\mathbf{H}}_{k_c})=Q. 
\end{array}
\end{equation}
Finally, to deal with the low-rank constraint in (\ref{prb:LS estimation l1 norm G}) and (\ref{prb:LS estimation l1 norm H}), it can be seen that it actually corresponds to a Riemannian manifold space, and thus the MO method \cite{sparse_property_channel_estimation_2022}\cite{LiuYue_ICC_2022} can be applied, where the optimization variable is iteratively updated in the direction of the Riemannian gradient and then is retracted back into the complex fixed-rank MO to make the result satisfy the low-rank constraint. 


The crucial step is to derive the Riemannian gradient, which can be deduced from the classic conjugate gradient in the Euclidean space. By using some properties in matrix derivation such as $\mathrm{d}(f)=\mathrm{Tr}\big(\nabla_{\mathbf{X}^*}f\mathrm{d}(\mathbf{X}^H)\big)$, 
$\mathrm{d}(\|\mathbf{r}-\mathbf{DXb}\|^2)=\mathrm{Tr}\Big(\big(-\mathbf{D}^H\mathbf{r}\mathbf{b}^H+\mathbf{D}^H\mathbf{D}\mathbf{X}\mathbf{b}\mathbf{b}^H\big)\mathrm{d}(\mathbf{X}^H)\Big)$
and $\mathrm{d}(\|\ \boldsymbol{\lambda}_{\widehat{\mathbf{G}}}\|_1)=\frac{1}{2}\mathrm{Tr}\Big(\big(\mathbf{A}_{\mathrm{BS}} \mathbf{Y}_{\widehat{\mathbf{G}}}\mathbf{A}_{\mathrm{I}}^{H}\big)\mathrm{d}(\mathbf{X}^H)\Big)$, the Euclidean conjugate gradient of the objective function in (\ref{prb:LS estimation l1 norm G}) can be found to be given by
\begin{equation}\label{Euclidean conjugate gradient G}
\begin{array}{rl}
    \nabla_{{\widehat{\mathbf{G}}^*}}f_{\widehat{\mathbf{G}}}=&\sum_{t=1}^{{T_\mathbf{G}}}\Big(-\sqrt{1-\epsilon}\mathbf{r}[t]\mathbf{s}^H[t]{\widehat{\mathbf{H}}^H_{k_c}\boldsymbol{\Phi}_{2}^{H}[t] \mathbf{L}^{H} \boldsymbol{\Phi}_{1}^{H}[t]}+(1-\epsilon){\widehat{\mathbf{G}}}\boldsymbol{\Phi}_{1}[t] \mathbf{L}\boldsymbol{\Phi}_{2}[t]\\ 
    &\times{\widehat{\mathbf{H}}_{k_c}} \mathbf{s}[t]\mathbf{s}^H[t]{\widehat{\mathbf{H}}^H_{k_c}\boldsymbol{\Phi}_{2}^{H}[t] \mathbf{L}^{H} \boldsymbol{\Phi}_{1}^{H}[t]}\Big)+\frac{\upsilon_{\widehat{\mathbf{G}}}}{2} \mathbf{A}_{\mathrm{BS}} \mathbf{Y}_{\widehat{\mathbf{G}}} \mathbf{A}_{\mathrm{I}}^{H},
\end{array}
\end{equation}
where $\mathbf{Y}_{\widehat{\mathbf{G}}}$ is computed as $\big[\mathbf{Y}_{\widehat{\mathbf{G}}}\big]_{ij}=\big[\mathbf{A}_{\mathrm{BS}}^H{\widehat{\mathbf{G}}}\mathbf{A}_{\mathrm{I}}\big]_{ij}\big/\big|\mathbf{A}_{\mathrm{BS}}^H{\widehat{\mathbf{G}}}\mathbf{A}_{\mathrm{I}}\big]_{ij}\big|$. Similarly, the Euclidean conjugate gradient of the objective function in (\ref{prb:LS estimation l1 norm H}) is given by 
\begin{equation}\label{Euclidean conjugate gradient H}
\begin{array}{rl}
     \nabla_{{\widehat{\mathbf{H}}_{k_c}}^*}f_{\widehat{\mathbf{H}}_{k_c}}=& \sum_{t=1}^{{T_\mathbf{G}}}\Big(-\sqrt{1-\epsilon}\boldsymbol{\Phi}_{2}^{H}[t] \mathbf{L}^{H} \boldsymbol{\Phi}_{1}^{H}[t]  \widehat{\mathbf{G}}^{H}\mathbf{r}[t]\mathbf{s}^{H}[t]+(1-\epsilon)\boldsymbol{\Phi}_{2}^{H}[t] \mathbf{L}^{H}\\
     &\times\boldsymbol{\Phi}_{1}^{H}[t]\widehat{\mathbf{G}}^{H} \widehat{\mathbf{G}}\boldsymbol{\Phi}_{1}[t] \mathbf{L}\boldsymbol{\Phi}_{2}[t] \widehat{\mathbf{H}}_{k_c}\mathbf{s}[t] \mathbf{s}^{H}[t] \Big)+\frac{\upsilon_{\widehat{\mathbf{H}}_{k_c}}}{2} \mathbf{A}_{\mathrm{I}} \mathbf{Y}_{\widehat{\mathbf{H}}_{k_c}} \mathbf{A}_{\mathrm{UE}}^{H},
\end{array}
\end{equation}
where  $\mathbf{Y}_{\widehat{\mathbf{H}}_{k_c}}$ is denoted as $\big[\mathbf{Y}_{\widehat{\mathbf{H}}_{k_c}}\big]_{ij}=\big[\mathbf{A}_{\mathrm{I}}^H{\widehat{\mathbf{H}}_{k_c}}\mathbf{A}_{\mathrm{UE}}\big]_{ij}\big/\big|\mathbf{A}_{\mathrm{I}}^H{\widehat{\mathbf{H}}_{k_c}}\mathbf{A}_{\mathrm{UE}}\big]_{ij}\big|$.

With the derived Euclidean conjugate gradient, we can project it onto the tangent space to obtain the Riemannian gradient. Then, by iteratively updating the corresponding variable with the Armijo backtracking step \cite{Armij} and retracting it back to the complex fixed-rank manifold, $\widehat{\mathbf{G}}$ and $\widehat{\mathbf{H}}_{k_c}$ can be  alternatively estimated with the other one fixed. The overall algorithm is referred to as the MO-CE algorithm and is summarized in \textbf{Algorithm 1}.
\begin{algorithm}[t]
	\caption{The MO-CE algorithm in the large timescale estimation}
	\begin{algorithmic}[1]
	\STATE Randomly initialize $\mathbf{\widehat{\mathbf{G}}}^{(0)},\mathbf{\widehat{H}}_{k_c}^{(0)}$. Set $i=0$.
\REPEAT
    \STATE Calculate $\nabla_{{\widehat{\mathbf{G}}^*}}f_{\widehat{\mathbf{G}}}$ according to (\ref{Euclidean conjugate gradient G}) with $\widehat{\mathbf{G}}^{(i)}$ and $\mathbf{\widehat{H}}_{k_c}^{(i)}$.
    \STATE Update $\widehat{\mathbf{G}}^{(i+1)}$ via the MO method for given $\mathbf{\widehat{H}}_{k_c}^{(i)}$.
    \STATE Calculate $\nabla_{{\widehat{\mathbf{H}}_{k_c}}^*}f_{\widehat{\mathbf{H}}_{k_c}}$ according to (\ref{Euclidean conjugate gradient H}) with $\widehat{\mathbf{G}}^{(i+1)}$ and $\mathbf{\widehat{H}}_{k_c}^{(i)}$.
    \STATE Update $\widehat{\mathbf{H}}_{k_c}^{(i+1)}$ via the MO method for given ${\widehat{\mathbf{G}}}^{(i+1)}$.
    \STATE $i\leftarrow i+1$.
\UNTIL the stopping condition is satisfied.
	\end{algorithmic}
\end{algorithm}

\subsection{Estimation of $\mathbf{H}_k$ for All UEs in the Small Timescale}\label{subsec:estimation of H}
As the BIOS-BS channel ${\mathbf{G}}$ varies much slower than the UE-BIOS channels, in the small timescale, the BS can separately estimate them for all UEs over consecutive time based on its estimated BIOS-BS channel $\widehat{\mathbf{G}}$ in the large timescale. Without loss of generality, we focus on the estimation of $\mathbf{H}_k$ with $T_\mathrm{H}$ uplink pilot vectors sent by the $k$-th UE. Considering the channel sparsity, the estimation problem of $\mathbf{H}_{k}$ with the LS objective similar to that in \eqref{prb:LS estimation l1 norm H} is formulated as follows
\begin{equation}\label{prb:LS estimation l1 norm H 2}
\begin{array}{cl}
\underset{\widehat{\mathbf{H}}_{k}}{\operatorname{minimize}} & g_{\widehat{\mathbf{H}}_{k}}=\sum_{t=1}^{T_\mathrm{H}}
\big\|\mathbf{r}[t]-\widehat{\mathbf{G}}\widehat{\boldsymbol{\Phi}}_{\mathrm{\mu}(k)}[t]\widehat{\mathbf{H}}_{k}\mathbf{s}[t]\big\|^{2} + \upsilon_{\widehat{\mathbf{H}}_{k}}\big\|\boldsymbol{\lambda}_{\widehat{\mathbf{H}}_{k}}\big\|_{1}\\
\text { subject to } &  \operatorname{rank}(\widehat{\mathbf{H}}_{k})=Q, 
\end{array}
\end{equation}
where $\mathbf{r}[t]$ is given in \eqref{eqn:received signal}, $\mathrm{\mu}(k)=\mathrm{fle}$ when $k=1,\dots,K_{\mathrm{fle}}$, and $\mathrm{\mu}(k)=\mathrm{fra}$ when $k=K_{\mathrm{fle}}+1,\dots,K$. It is worth noting that \eqref{prb:LS estimation l1 norm H 2} is suitable for both  UE$_{\mathrm{fra}}$s and UE$_{\mathrm{fle}}$s. As (\ref{prb:LS estimation l1 norm H 2}) has the same form as \eqref{prb:LS estimation l1 norm H}, it can also be solved by the MO method. It can be derived that the Euclidean gradients $\nabla_{{\widehat{\mathbf{H}}_{k}}^*}g_{\widehat{\mathbf{H}}_{k}}$ is given by 
\begin{equation}\label{Euclidean conjugate gradient H 2}
\begin{array}{ll}
     \nabla_{{\widehat{\mathbf{H}}_{k}}^*}g_{\widehat{\mathbf{H}}_{k}}=&\sum_{t=1}^{T_{\mathrm{H}}}\Big(-\widehat{\boldsymbol{\Phi}}^H_{\mathrm{\mu}(k)}[t]  \widehat{\mathbf{G}}^{H}\mathbf{r}[t]\mathbf{s}^{H}[t] +\widehat{\boldsymbol{\Phi}}^H_{\mathrm{\mu}(k)}[t]\widehat{\mathbf{G}}^{H} \widehat{\mathbf{G}}\widehat{\boldsymbol{\Phi}}_{\mathrm{\mu}(k)}[t]\widehat{\mathbf{H}}_{k}\mathbf{s}[t]\mathbf{s}^{H}[t] \Big) \\

&+\frac{\upsilon_{\widehat{\mathbf{H}}_{k}}}{2} \mathbf{A}_{\mathrm{I}} \mathbf{Y}_{\widehat{\mathbf{H}}_{ k}} \mathbf{A}_{\mathrm{UE}}^{H}.
\end{array}
\end{equation}
The Riemannian gradient can then be obtained by projecting the Euclidean gradient onto the tangent space. By updating the variable iteratively via the MO method until convergence,  $\widehat{\mathbf{H}}_{k}$ can be finally obtained. 

\subsection{Analysis of  Training Overhead}
In this subsection, we analyze the training overhead of the proposed HTT-MO scheme in terms of the required number of pilot vectors and that of the conventional LS scheme. By recalling Fig. \ref{two step strategy}, we can see that the overall training overhead of the HTT-MO scheme is given by $T_{\mathrm{tot}}=T_{\mathrm{G}}+\tau KT_{\mathrm{H}}$, where $\tau$ is the ratio of the length of a large timescale to that of a small timescale. It is worth noting in the large timescale both ${\mathbf{G}}$ and ${\mathbf{H}}_{k_c}$ need to be estimated while in the small timescale only ${\mathbf{H}}_{k}$ needs to be estimated for each UE. Furthermore, as the estimation result $\widehat{\mathbf{G}}$ in the large timescale must be used for the estimation of ${\mathbf{H}}_{k}$ in the small timescale, the estimation accuracy of $\widehat{\mathbf{G}}$ must be higher. Thus, $T_{\mathrm{G}}$ should be larger than $T_{\mathrm{H}}$.

For comparison, we consider the traditional LS channel estimation scheme for the proposed BIOS-assisted system. As it does not exploit the heterogeneous, two-timescale and sparsity properties of channels, the traditional LS channel estimation scheme has to estimate the high-dimensional cascaded channels for all UE$_\mathrm{fle}$s and UE$_\mathrm{fra}$s in each small timescale, and thus results in prohibitive pilot overhead. To analyze the number of required pilot vectors in this traditional channel estimation scheme, we rewrite the equivalent baseband received signal of the BIOS-assisted system in (\ref{eqn:received signal fle}) and (\ref{eqn:received signal fra}) as follows
\begin{equation}\label{eqn:received signal fle2}
 \begin{array}{rll}
\mathbf{r}[t]
&=&\sqrt{\epsilon}(\mathbf{s}^T[t]\otimes\mathbf{I}_{{N}_\mathrm{BS}}){\mathbf{J}}_{\mathrm{fle},k}{\boldsymbol{\phi}}_1[t] +\mathbf{z}[t]\\
&=&\sqrt{\epsilon}\big({\boldsymbol{\phi}}_1[t]^T\otimes(\mathbf{s}^T[t]\otimes\mathbf{I}_{{N}_\mathrm{BS}})\big)\mathrm{vec}({\mathbf{J}}_{\mathrm{fle},k}) +\mathbf{z}[t],
\end{array}
\end{equation}
\begin{equation}\label{eqn:received signal fra2}
\begin{array}{rll}
\mathbf{r}[t]&=&\sqrt{1-\epsilon}(\mathbf{s}^T[t]\otimes\mathbf{I}_{{N}_\mathrm{BS}}){\mathbf{J}}_{\mathrm{fra},k}\mathrm{vec}(\mathbf{\Phi}_1[t]\mathbf{L}\mathbf{\Phi}_2[t]) +\mathbf{z}[t]\\
&=&\sqrt{1-\epsilon}(\mathrm{vec}(\mathbf{\Phi}_1[t]\mathbf{L}\mathbf{\Phi}_2[t]))^T\otimes(\mathbf{s}^T[t]\otimes\mathbf{I}_{{N}_\mathrm{BS}})\big)\mathrm{vec}({\mathbf{J}}_{\mathrm{fra},k}) +\mathbf{z}[t],
\end{array}
\end{equation}
where (\ref{eqn:received signal fle2}) corresponds to UE$_{\mathrm{fle}}$s, and (\ref{eqn:received signal fra2}) corresponds to UE$_{\mathrm{fra}}$s. In this situation, the traditional LS channel estimation scheme needs to estimate the high-dimensional cascaded channels, i.e., ${\mathbf{J}}_{\mathrm{fle},k}{=}\mathbf{H}^T_{k}\odot\mathbf{G}$ for UE$_{\mathrm{fle}}$s and ${\mathbf{J}}_{\mathrm{fra},k}{=}\mathbf{H}^T_{k}\otimes\mathbf{G}$ for UE$_{\mathrm{fra}}$s, and the corresponding number of required pilot vectors is at least $MN_{\mathrm{UE}}$ and $M^2N_{\mathrm{UE}}$, respectively. As the cascaded channels of all UEs need to be updated in each small timescale, the total number of pilot vectors in the LS estimation scheme in a large timescale is at least $\tau \big(K_{\mathrm{fle}}MN_{\mathrm{UE}}+K_{\mathrm{fra}}M^2N_{\mathrm{UE}}\big)$, which is more than $2.3\times 10^5$ 
if setting $\tau=4$, $K_{\mathrm{fle}}=2$, $K_{\mathrm{fra}}=3$, $M=49$ and $N_{\mathrm{UE}}=8$. Such amount of overhead cannot be afforded in practical systems. In contrast, simulation results in Section \ref{sec:Simulation Results} will show that in the BIOS-assisted system with the same setup, only about two thousands of pilot vectors are required by the HTT-MO scheme in a large timescale.

\section{WMMSE-CD Beamforming Optimization}\label{sec:beamforming}
With the estimated CSI by the HTT-MO scheme in Section \ref{sec:MO-EST}, in this section, we focus on the multi-user downlink beamforming optimization, and propose the WMMSE-CD scheme to maximize sum data rate of all UEs on both sides of the BIOS. 

\subsection{Problem Formulation}\label{subsec:Problem formulation}
Assuming that the uplink and downlink channels are reciprocal, the estimated CSI of the uplink BIOS-BS and UE-BIOS channels can be utilized in the downlink beamforming optimization. Assuming that the BS sends $N_{\mathrm{s}}$ data streams to each UE, the received signal at the $k$-th UE, $\mathbf{y}_{k}\in\mathbb{C}^{N_{\mathrm{UE}}\times1}$, can be represented by
\begin{equation}\label{eqn:downlink signal}
    \mathbf{y}_{k}={\mathbf{H}}_{k}^{H}{\widehat{\mathbf{\Phi}}_{\mathrm{d},\mathrm{\mu}(k)}}\mathbf{G}^H\mathbf{F}\mathbf{s}_{\mathrm{d}}+\mathbf{n}_{k},
\end{equation}
where $\mathbf{F} = [\mathbf{F}_1,\dots,\mathbf{F}_K]\in\mathbb{C}^{N_{\mathrm{BS}}\times N_{\mathrm{s}}K}$ is the BS precoder,  $\mathbf{s}_{\mathrm{d}}=[\mathbf{s}_{\mathrm{d},1}^H,\dots,\mathbf{s}_{\mathrm{d},K}^H]^H\in\mathbb{C}^{N_{\mathrm{s}}K\times 1}$ is the symbol vector with $\mathbb{E}\{\mathbf{s}_{\mathrm{d}}\mathbf{s}_{\mathrm{d}}^H\}=\mathbf{I}_{N_{\mathrm{s}}K}$, and  $\mathbf{n}_{k}\sim\mathcal{CN}(0, \sigma_{\mathrm{d}}^2\mathbf{I}_{N_{\mathrm{UE}}})\in\mathbb{C}^{N_{\mathrm{UE}}\times 1}$ is the noise at the $k$-th UE. Then, the effective data rate (per Hertz) of the $k$-th UE can be expressed as 
\begin{equation}\label{eqn:efficient SE}
\begin{array}{l}
    {R}_k=\left(1-\frac{T_{\mathrm{tot}}}{\Upsilon_{\mathrm{large}}}\right){{\rm{log}}\left|\mathbf{I}_{N_{\mathrm{s}}}+\mathbf{F}_{k}^H\mathbf{H}^H_{\mathrm{e},k}\boldsymbol{\Lambda}^{-1}_k\mathbf{H}_{\mathrm{e},k}\mathbf{F}_{k}\right|},
\end{array}
\end{equation}
where $\mathbf{H}_{\mathrm{e},k}={\mathbf{H}}_{k}^{H}{\widehat{\mathbf{\Phi}}_{\mathrm{d},\mathrm{\mu}(k)}}\mathbf{G}^H\in\mathbb{C}^{N_{\mathrm{UE}}\times N_{\mathrm{BS}}}$ denotes the efficient channel matrix from the BS to the $k$-th UE, and $\boldsymbol{\Lambda}_{k}=\sigma_{\mathrm{d}}^{2} \mathbf{I}_{N_\mathrm{UE}}+\sum_{i \neq k}^{K} {\mathbf{H}_{\mathrm{e},k}} \mathbf{F}_{i} \mathbf{F}_{i}^{H}{\mathbf{H}^H_{\mathrm{e},k}}\in\mathbb{C}^{N_{\mathrm{UE}}\times N_{\mathrm{UE}}}$ denotes covariance of the noise plus multiuser interference at the $k$-th UE. $\Upsilon_{\mathrm{large}}$ is the length of a large timescale in terms of the number of symbols within this period as shown in Fig. \ref{two step strategy}, and ${T_{\mathrm{tot}}}$ denotes the total training overhead in terms of the number of pilot symbols. 

According to \cite{WMMSE}, the sum rate maximization (SRM) problem can be solved via an equivalent WMMSE problem. It turns out that this optimization approach can also be applied for solving the joint BS-BIOS beamforming optimization problem with the SRM objective, and the equivalent WMMSE problem is given by
%
\begin{equation}\label{prb:WMMSE}
\begin{array}{cl}
\displaystyle{\minimize_{{{\mathbf{F}}, \mathbf{\Phi}_{\mathrm{d},1}, \mathbf{\Phi}_{\mathrm{d},2},\mathbf{W}_k,\mathbf{\Psi}_k}}} & {\sum_{k=1}^K\mathrm{Tr}(\mathbf{\Psi}_k\mathbf{E}_k)-\mathrm{log}|\mathbf{\Psi}_k| }  \\
\mathrm{subject \; to} & {\rm{Tr}}(\mathbf{F}\mathbf{F}^H)\le{1},\\ 
\quad & |[\mathbf{\Phi}_{\mathrm{d},1}]_{m_1m_1}| = 1,\ |[\mathbf{\Phi}_{\mathrm{d},2}]_{m_2m_2}| = 1,\quad \forall m_1, m_2,
\end{array}
\end{equation}
where the first constraint represents the normalized transmit power constraint at the BS, and the second one corresponds to the constant modulus constraint of the BIOS passive scattering elements. $\mathbf{E}_k=\mathbb{E}\left[\left(\mathbf{s}_{\mathrm{d},k}-\mathbf{W}^H_k\mathbf{y}_k\right)\left(\mathbf{s}_{\mathrm{d},k}-\mathbf{W}^H_k\mathbf{y}_k\right)^H\right]\in\mathbb{C}^{N_{\mathrm{s}}\times N_{\mathrm{s}}}$ is the mean square error (MSE) matrix of the $k$-th UE, with $\mathbf{W}_k\in\mathbb{C}^{N_{\mathrm{UE}}\times N_{\mathrm{s}}}$ denoting the combining matrix. $\mathbf{\Psi}_k\in\mathbb{C}^{N_{\mathrm{s}}\times N_{\mathrm{s}}}$ is an auxiliary variable to establish the equivalence between the SRM problem and the WMMSE problem. 

To deal with the multivariate optimization difficulty, the alternating minimization method is applied. Firstly, $\mathbf{W}_k$ and $\mathbf{\Psi}_k$ are optimized while keeping other variables fixed, which can be shown to have the following closed-form solutions
\begin{equation}\label{closed-form solution of W and Psi}
    \begin{array}{rll}
         \mathbf{W}_k^\star&=&\Big(\mathbf{\Lambda}_k+\mathbf{H}_{\mathrm{e},k}\mathbf{F}_k\mathbf{F}_k^H\mathbf{H}^H_{\mathrm{e},k}\Big)^{-1}\mathbf{H}_{\mathrm{e},k}\mathbf{F}_k,\\ \mathbf{\Psi}_k^\star&=&\left(\mathbf{E}_k\right)^{-1}.
    \end{array}
\end{equation} 
Then, to optimize the BS precoder and the BIOS passive beamformers while keeping $\mathbf{W}_k$ and $\mathbf{\Psi}_k$ fixed, by expanding the MSE matrix in \eqref{prb:WMMSE} and removing the terms that are not related to the optimization variables, problem (\ref{prb:WMMSE}) can be simplified as 
\begin{equation}\label{prb:WMMSE3}
\begin{array}{cl}
\displaystyle{\minimize_{{{\mathbf{F}}, \mathbf{\Phi}_{\mathrm{d},1}, \mathbf{\Phi}_{\mathrm{d},2}}}} & { f=\mathrm{Tr}\left(\mathbf{\Psi}\mathbf{W}^H\mathbf{H}_\mathrm{e}\mathbf{F}\mathbf{F}^H\mathbf{H}_\mathrm{e}^H\mathbf{W}-\mathbf{\Psi}\mathbf{W}^H\mathbf{H}_\mathrm{e}\mathbf{F}-\mathbf{\Psi}\mathbf{F}^H\mathbf{H}_\mathrm{e}^H\mathbf{W}\right)} \\
\mathrm{subject \; to} & {\rm{Tr}}(\mathbf{F}\mathbf{F}^H)\le{1},\\ 
\quad & |[\mathbf{\Phi}_{\mathrm{d},1}]_{m_1m_1}| = 1,\ |[\mathbf{\Phi}_{\mathrm{d},2}]_{m_2m_2}| = 1,\quad \forall m_1, m_2.
\end{array}
\end{equation} 
where  $\mathbf{\Psi}=\mathrm{blkdiag}\left(\mathbf{\Psi}_1,\dots,\mathbf{\Psi}_K\right)$, $\mathbf{W}=\mathrm{blkdiag}\left(\mathbf{W}_1,\dots,\mathbf{W}_K\right)$, $\mathbf{E}=\mathrm{blkdiag}\left(\mathbf{E}_1,\dots,\mathbf{E}_K\right)$ and $\mathbf{H}_\mathrm{e} = \left(\mathbf{H}_{\mathrm{e},1}^T,\dots,\mathbf{H}_{\mathrm{e},K}^T\right)^T$. 

\subsection{Optimization of the BS Precoder}\label{subsec:Optimization of the BS Precoder}
To optimize the BS precoder in \eqref{prb:WMMSE3} while keeping the BIOS passive beamformers  $\mathbf{\Phi}_{\mathrm{d,1}}$ and $\mathbf{\Phi}_{\mathrm{d,2}}$ fixed, it can be shown that the BS precoder $\mathbf{F}$ has the following closed-form solution according to the Karush-Kuhn-Tucker (KKT) conditions \cite{WMMSE} 
\begin{equation}\label{eqn:DPsolution}
    \mathbf{F}^{\star} =\zeta\tilde{\mathbf{F}}^{-1} \mathbf{H}_\mathrm{e}^H\mathbf{W}\mathbf{\Psi},
\end{equation}
where $\tilde{\mathbf{F}} = \mathbf{H}_\mathrm{e}^H\mathbf{W}\mathbf{\Psi}\mathbf{W}^H\mathbf{H}_\mathrm{e}+{\sigma_{\mathrm{d}}^2}\mathrm{Tr}\left(\mathbf{\Psi}\mathbf{W}^H\mathbf{W}\right)\mathbf{I}_{N_{\mathrm{BS}}}$ and $\zeta=\|\tilde{\mathbf{F}}^{-1} \mathbf{H}_\mathrm{e}^H\mathbf{W}\mathbf{\Psi}\|^{-1}_F$.

\subsection{Optimization of the IOS$_1$}\label{Optimization of the IOS1}
When optimizing the passive beamformer  $\mathbf{\Phi}_{\mathrm{d,1}}$ while keeping other variables fixed in \eqref{prb:WMMSE3}, the difficulty is the non-convex constant modulus constraint of the scattering elements. One way to solve this difficulty is to apply the CD algorithm \cite{CDA}, and alternatively optimize each element of $\mathbf{\Phi}_{\mathrm{d},1}$. By defining the equivalent IOS$_1$-UEs channel as 
\begin{equation*}
    \mathbf{H}^H_{\mathbf{\Phi}}=\left[(\sqrt{\epsilon}\mathbf{H}^H_{\mathrm{1}})^T, \dots,(\sqrt{\epsilon}\mathbf{H}^H_{K_\mathrm{fle}})^T, (\sqrt{1-\epsilon}\mathbf{H}^H_{K_\mathrm{fle}+1}\mathbf{\Phi}_{\mathrm{d,2}} \mathbf{L}^H)^T, \dots,(\sqrt{1-\epsilon}\mathbf{H}^H_{K}\mathbf{\Phi}_{\mathrm{d,2}}\mathbf{L}^H)^T\right]^T,
\end{equation*}
the objective function $f$ in \eqref{prb:WMMSE3} can be represented as 
\begin{equation}\label{eqn:f_rewrite_Phi}
\begin{array}{rll}
f&=&\mathrm{Tr}\big(\mathbf{\Psi}\mathbf{W}^H\mathbf{H}^H_\mathbf{\Phi}\mathbf{\Phi}_{\mathrm{d,1}}\mathbf{G}^H\mathbf{F}{\mathbf{F}}^H\mathbf{G}\mathbf{\Phi}_{\mathrm{d,1}}^H\mathbf{H}_{\mathbf{\Phi}}\mathbf{W}-\mathbf{\Psi}\mathbf{W}^H\mathbf{H}^H_{\mathbf{\Phi}}\mathbf{\Phi}_{\mathrm{d,1}}\mathbf{G}^H\mathbf{F}\\
& &-\mathbf{\Psi}{\mathbf{F}}^H\mathbf{G}\mathbf{\Phi}_{\mathrm{d,1}}^H\mathbf{H}_{\mathbf{\Phi}}\mathbf{W}\big)\\
&\overset{(\mathrm{a})}{=}&\boldsymbol{\phi}_{\mathrm{d,1}}^H\mathbf{\Xi}\boldsymbol{\phi}_{\mathrm{d,1}}-\boldsymbol{\rho}^H\boldsymbol{\phi}_{\mathrm{d,1}}-\boldsymbol{\phi}_{\mathrm{d,1}}^H\boldsymbol{\rho},
\end {array}
\end{equation}
where $\mathbf{\Xi}=\big(\mathbf{H}_{\mathbf{\Phi}}\mathbf{W}\mathbf{\Psi}\mathbf{W}^H\mathbf{H}^H_\mathbf{\Phi}\big)\circ\big(\mathbf{G}^H\mathbf{F}{\mathbf{F}}^H\mathbf{G}\big)^T$,  $\boldsymbol{\rho}=\mathrm{diag}\big(\mathbf{H}_{\mathbf{\Phi}}\mathbf{W}\mathbf{\Psi}{\mathbf{F}}^H\mathbf{G}\big)$ and $\boldsymbol{\phi}_{\mathrm{d,1}}=\mathrm{diag}(\boldsymbol{\Phi}_{\mathrm{d,1}})$. The equality $(\mathrm{a})$ follows from the facts that $\mathrm{Tr}(\mathbf{AB})=\mathrm{Tr}(\mathbf{BA})$, $\mathrm{Tr}(\mathbf{A+B})=\mathrm{Tr}(\mathbf{A})+\mathrm{Tr}(\mathbf{B})$, ${\rm{Tr}}(\mathbf{C}^H\mathbf{A}\mathbf{C}\mathbf{B})=\mathbf{c}^H(\mathbf{A}\circ\mathbf{B}^T)\mathbf{c}$ and ${\rm{Tr}}(\mathbf{A}\mathbf{C})=\mathbf{a}^T\mathbf{c}$ for  arbitrary matrices $\mathbf{A}$, $\mathbf{B}$ and diagonal matrix $\mathbf{C}$, with $\mathbf{a}=\rm{diag}(\mathbf{A})$ and $\mathbf{c}=\rm{diag}(\mathbf{C})$. Without loss of generality, we assume that the element to be optimized in the current
iteration of the CD algorithm is $\left[\boldsymbol{\phi}_{\mathrm{d,1}}\right]_{m_1}$ with other elements of $\boldsymbol{\phi}_{\mathrm{d,1}}$ fixed. Then, by omitting the constant terms irrelevant to the optimization of $\left[\boldsymbol{\phi}_{\mathrm{d,1}}\right]_{m_1}$, the objective function $f$ can be simplified as  
\begin{equation}\label{eqn:f12}
    f= 2\mathcal{R}\{(\sum_{m_1'\neq{m_1}}[\mathbf{\Xi}]_{{m_1m_1'}}[\boldsymbol{\phi}_{\mathrm{d,1}}]_{m_1'}-[\boldsymbol{\rho}]_{m_1})[\boldsymbol{\phi}_{\mathrm{d,1}}]^*_{m_1}\}.
\end{equation}
It can be shown that the optimal solution of $[\boldsymbol{\phi}_{\mathrm{d,1}}]_{m_1}$ with other elements fixed is given by
\begin{equation}\label{CD solution}
[\boldsymbol{\phi}_{\mathrm{d,1}}]_{m_1}= -\frac{\sum_{m_1'\neq{m_1}}[\mathbf{\Xi}]_{{m_1m_1'}}[\boldsymbol{\phi}_{\mathrm{d,1}}]_{m_1'}-[\boldsymbol{\rho}]_{m_1}}{\big|\sum_{m_1'\neq{m_1}}[\mathbf{\Xi}]_{{m_1m_1'}}[\boldsymbol{\phi}_{\mathrm{d,1}}]_{m_1'}-[\boldsymbol{\rho}]_{m_1}\big|}.
\end{equation}
\subsection{Optimization of the IOS$_2$}\label{Optimization of the IOS2}
As the performance of UE$_{\mathrm{fle}}$ is not related to $\boldsymbol{\Phi}_{\mathrm{d,2}}$, we can divide $f$ in \eqref{prb:WMMSE3} into two terms, with one corresponding to the UE$_{\mathrm{fra}}$s, which is related to $\boldsymbol{\Phi}_{\mathrm{d,2}}$, and the other to the UE$_{\mathrm{fle}}$s, which can be taken as a constant when optimizing $\boldsymbol{\Phi}_{\mathrm{d,2}}$. By defining  $\mathbf{H}^H_{\mathbf{\Phi},\mathrm{fra}}=\Big(\big(\sqrt{1-\epsilon}\mathbf{H}^H_{K_\mathrm{fle}+1}\big)^T, \dots,$\\$\big(\sqrt{1-\epsilon}\mathbf{H}^H_{K}\big)^T\Big)^T$, $f$ can be further reformed as
\begin{equation}\label{eqn:f_rewrite_Theta}
\begin{array}{cl}
f
&=\mathrm{Tr}(\mathbf{\Psi}_{\mathrm{fra}}\mathbf{W}^H_{\mathrm{fra}}\mathbf{H}^H_{\mathbf{\Phi},\mathrm{fra}}\boldsymbol{\Phi}_{\mathrm{d,2}}\mathbf{L}^H\boldsymbol{\Phi}_{\mathrm{d,1}} \mathbf{G}^H\mathbf{F}\mathbf{F}^H\mathbf{G}\boldsymbol{\Phi}_{\mathrm{d,1}}^H\mathbf{L}\boldsymbol{\Phi}_{\mathrm{d,2}}^{H}\mathbf{H}_{\mathbf{\Phi},\mathrm{fra}}\mathbf{W}_{\mathrm{fra}})-\mathrm{Tr}(\mathbf{\Psi}_{\mathrm{fra}}\times\\
&\quad\mathbf{W}^H_{\mathrm{fra}}\mathbf{H}^H_{\mathbf{\Phi},\mathrm{fra}}\boldsymbol{\Phi}_{\mathrm{d,2}}\mathbf{L}^H\boldsymbol{\Phi}_{\mathrm{d,1}} \mathbf{G}^H\mathbf{F}_{\mathrm{fra}})-\mathrm{Tr}(\mathbf{\Psi}_{\mathrm{fra}}\mathbf{F}_{\mathrm{fra}}^{H} \mathbf{G}\boldsymbol{\Phi}_{\mathrm{d,1}}^{H} \mathbf{L} \boldsymbol{\Phi}_{\mathrm{d,2}}^H \mathbf{H}_{\mathbf{\Phi},\mathrm{fra}}\mathbf{W}_{\mathrm{fra}}) + \mathrm{const}\\
&=\boldsymbol{\phi}_{\mathrm{d,2}}^H\mathbf{\Xi}_{\mathrm{fra}}\boldsymbol{\phi}_{\mathrm{d,2}}-\boldsymbol{\rho}_{\mathrm{fra}}^H\boldsymbol{\phi}_{\mathrm{d,2}}-\boldsymbol{\phi}_{\mathrm{d,2}}^H\boldsymbol{\rho}_{\mathrm{fra}}+ \mathrm{const},
\end{array}
\end{equation}
where $\mathbf{\Psi}_{\mathrm{fra}} = \mathrm{blkdiag}(\mathbf{\Psi}_{K_{\mathrm{fle}}+1},\cdots, \mathbf{\Psi}_{K})$,  $\mathbf{W}_{\mathrm{fra}}=\mathrm{blkdiag}(\mathbf{W}_{K_{\mathrm{fle}}+1},\cdots, \mathbf{W}_{K})$, $\mathbf{F}_{\mathrm{fra}}=[\mathbf{F}_{K_{\mathrm{fle}}+1},$\\$\cdots, \mathbf{F}_{K}]$, $\mathbf{\Xi}_{\mathrm{fra}}=\big(\mathbf{H}_{\mathbf{\Phi},{\mathrm{fra}}}\mathbf{W}_{\mathrm{fra}}\mathbf{\Psi}_{\mathrm{fra}}\mathbf{W}_{\mathrm{fra}}^H\mathbf{H}^H_{\mathbf{\Phi},{\mathrm{fra}}}\big)\circ\big(\mathbf{L}^H\boldsymbol{\Phi}_{\mathrm{d,1}} \mathbf{G}^H\mathbf{F}\mathbf{F}^H\mathbf{G}\boldsymbol{\Phi}_{\mathrm{d,1}}^H\mathbf{L}\big)^T$,  $\boldsymbol{\rho}_{\mathrm{fra}}=\mathrm{diag}\big(\mathbf{H}_{\mathbf{\Phi},\mathrm{fra}}$\\$\times\mathbf{W}_{\mathrm{fra}}\mathbf{\Psi}_{\mathrm{fra}}\mathbf{F}_{\mathrm{fra}}^{H} \mathbf{G}\boldsymbol{\Phi}_{\mathrm{d,1}}^{H} \mathbf{L} \big)$ and $\boldsymbol{\phi}_{\mathrm{d,2}}=\mathrm{diag}(\boldsymbol{\Phi}_{\mathrm{d,2}})$. As (\ref{eqn:f_rewrite_Theta}) is similar to (\ref{eqn:f_rewrite_Phi}), $\boldsymbol{\Phi}_{\mathrm{d,2}}$ can also be optimized by the CD algorithm. 

Finally, the WMMSE-CD algorithm for the BIOS-assisted multi-user MIMO system can be accomplished by alternatively optimizing $\mathbf{\Psi}$, $\mathbf{W}$, $\mathbf{F}$, $\mathbf{\Phi}_{\mathrm{d,1}}$ and $\mathbf{\Phi}_{\mathrm{d,2}}$. As the objective function monotonically decreases after the optimization of each variable, the WMMSE-CD algorithm is guaranteed to converge to a locally optimal solution of problem (\ref{prb:WMMSE}). It is worth noting that although we use the true channel matrices $\mathbf{G}$ and $\mathbf{H}_k$ in the derivation of WMMSE-CD algorithm, the solutions of the variables in (\ref{closed-form solution of W and Psi}), (\ref{eqn:DPsolution}) and (\ref{CD solution}) still hold for the estimates $\widehat{\mathbf{G}}$ and $\widehat{\mathbf{H}}_k$ obtained by the proposed HTT-MO scheme, as there is only a scalar difference between $\widehat{\mathbf{G}}$($\widehat{\mathbf{H}}_k$) and ${\mathbf{G}}$(${\mathbf{H}}_k$) if there is no noise effect in the channel estimation.

\section{Simulation Results}\label{sec:Simulation Results}
\subsection{Simulation Setup}\label{subsec:Simulation Setup}
Consider a BIOS-assisted system where the IOS$_1$ is set in the simultaneous reflection and refraction mode with $\epsilon=0.5$. The number of UEs is set to $K=5$, with $2$ on the reflection side and $3$ on the refraction side, i.e., $K_{\mathrm{fle}}=2$, $K_{\mathrm{fra}}=3$. The number of ULA elements at the BS and UEs is  $N_{\mathrm{BS}}=N_{\mathrm{UE}}=8$, and that of the UPA elements of the BIOS is $M = M_{\mathrm{y}}\times M_{\mathrm{z}}=49$ with $M_{\mathrm{y}}=M_{\mathrm{z}}=7$. For all of the ULAs and the two UPAs, the distance between neighboring units is $\frac{1}{2}\lambda$, with $\lambda=0.03\mathrm{m}$ denoting the wavelength of carrier wave. For the BIOS, the distance between the two IOSs is $0.03\mathrm{m}$. For the channel model in (\ref{eqn:channel models}), the path number of both $\mathbf{G}$ and $\mathbf{H}$ is set to $5$, i.e., $P=Q=5$. Similar to that in \cite{ANM1}\cite{sparse_property_channel_estimation_2022}, the first path, i.e., $p=1$ ($q=1$), is set as the LoS path of $\mathbf{G}$ ($\mathbf{H}$) with its complex path gain distributed as $\alpha_1(\beta_{1,k})\sim\mathcal{CN}(0,1)$, while other paths are NLoS paths with the path gain distribution of $\mathcal{CN}(0,0.1)$. For the BS and UEs, the AoAs/AoDs are assumed to be uniformly distributed in $[0,\pi]$, while for the BIOS, the azimuth and elevation AoAs/AoDs are assumed to satisfy a uniform distribution in $[0,2\pi]$ and $[0,\frac{\pi}{4}]\cup[\frac{3\pi}{4},\pi]$, respectively\footnote{With this setup, we can reform (\ref{eqn:AyAz}) by setting $x^i_\mathrm{x} = -\frac{\sqrt{2}}{2} + (i-1)\frac{\sqrt{2}}{G_\mathrm{x}-1}$, $x^i_\mathrm{y} = -\frac{\sqrt{2}}{2} + (i-1)\frac{\sqrt{2}}{G_\mathrm{y}-1}$, as $\big(-\mathrm{sin}(\theta)\mathrm{sin}(\phi)\big)$, $\big(-\mathrm{sin}(\theta)\mathrm{cos}(\phi)\big)\in[-\frac{\sqrt{2}}{2},\frac{\sqrt{2}}{2}]$. In this situation, we still have $\mathbf{A}_{\mathrm{I}}^H\mathbf{A}_{\mathrm{I}}\approx\mathbf{I}_{M}$ due to the orthogonality between columns in $\mathbf{A}_{\mathrm{I}}$, and thus the Lemma \ref{lem: sparsity of angle} still holds.}. The elevation range $[0,\frac{\pi}{4}]$ is associated to the UE$_{\mathrm{fle}}$s, and $[\frac{3\pi}{4},\pi]$ is associated to the UE$_{\mathrm{fra}}$s. It is assumed that the BIOS-BS (UE-BIOS) channel is time invariant in a large (small) timescale, as shown in Fig. \ref{two step strategy}, and is independent for different large (small) timescale spans. The length of large (small) timescale is set to $\Upsilon_\mathrm{large}=10000$ ($\Upsilon_\mathrm{small}=2500$), corresponding to a channel coherence time of $20\mathrm{ms}$ ($5\mathrm{ms}$) with a $ 500\mathrm{kHz}$ transmission bandwidth. The uplink training pilot-to-noise-ratio (PNR) is defined as $\frac{1}{\sigma^2}$, while the downlink transmission signal-to-noise-ratio (SNR) is defined as $\frac{1}{\sigma_{\mathrm{d}}^2}$. For the HTT-MO scheme, all elements of $\mathbf{\Phi}_1[t]$ and  $\mathbf{\Phi}_2[t]$ are randomly initialized from the constant modulus set \big\{$x\big||x|=1$\big\} to ensure a random quasi-omnidirectional beam pattern \cite{quasi-omnidirectional beam_2018}, and the elements of the pilot vector $\mathbf{s}[t]$ are also randomly selected from this set. All simulation results are averaged over 200 realizations.

\subsection{Performance of the HTT-MO Channel Estimation Scheme }\label{subsec:Performance of G estimation in step one}

As mentioned in Section \ref{subsection:Recovery of G and H from the cascaded channel}, even when there is no noise effect, the obtained $\widehat{\mathbf{G}}$ and $\widehat{\mathbf{H}}_k$ of all UEs by the proposed HTT-MO scheme is still different from the real ${\mathbf{G}}$ and ${\mathbf{H}}_k$ by a coefficient, i.e., $\widehat{\mathbf{G}}=\frac{1}{a}\mathbf{G}$ and $\widehat{\mathbf{H}}_k=a\mathbf{H}_k$. Thus, it is not suitable to simply take the estimation MSE of ${\mathbf{G}}$ or ${\mathbf{H}}_k$ as a performance metric. Instead, we take the normalized MSE (NMSE) of $\mathbf{H}_k^T\otimes\mathbf{G}$ to evaluate the performance of channel estimation, which is defined as $\mathbb{E}\left\{\big\|\big(\mathbf{H}_k^T\otimes\mathbf{G}\big)-\big(\widehat{\mathbf{H}}_k^T\otimes\widehat{\mathbf{G}}\big)\big\|^2_{F}\big/\big\|\mathbf{H}_k^T\otimes\mathbf{G}\big\|^2_F \right\}$. 

\begin{figure}
	\centering
\includegraphics[height=7.6cm,width=14.9cm]{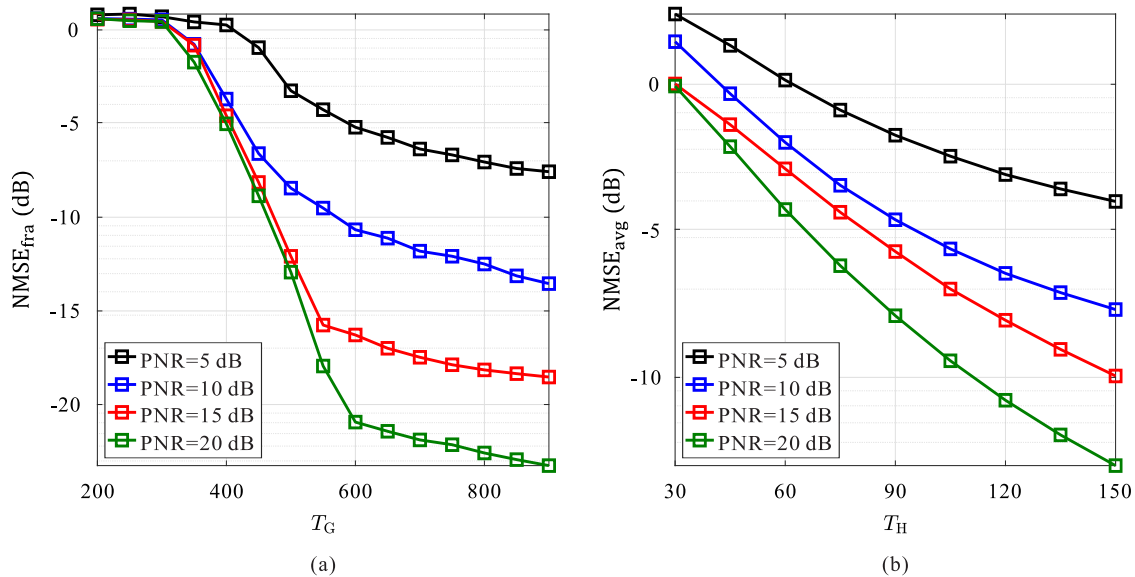}
	\caption{(a) $\mathrm{NMSE}_{\mathrm{fra}}$ versus $T_{\mathrm{G}}$ for the selected UE$_{\mathrm{fra}}$ in the large timescale. (b) $\mathrm{NMSE}_{\mathrm{avg}}$ versus $T_{\mathrm{H}}$ for all UEs with the $\widehat{\mathbf{G}}$ estimated in the large timescale and the $\widehat{\mathbf{H}}_k$ estimated in the small timescale.}	\label{fig:simulation1}
\end{figure}

We first evaluate the performance of the large timescale estimation step in the HTT-MO scheme. As introduced in Section \ref{sec:MO-EST}, a UE$_{\mathrm{fra}}$ is selected for the estimation of $\mathbf{G}$ in the large timescale estimation, and both $\mathbf{G}$ and its UE-BIOS channel are estimated via the MO-CE algorithm. Fig. \ref{fig:simulation1}(a) exhibits the NMSE performance of this selected UE$_{\mathrm{fra}}$, which is defined as $\mathrm{NMSE}_{\mathrm{fra}}$, for different values of PNR and training overhead $T_\mathrm{G}$. It can be seen that for small $T_{\mathrm{G}}$, e.g., $T_{\mathrm{G}}\leq300$, the proposed MO-CE algorithm suffers from a shortage of training overhead, resulting in an NMSE larger than $1$ regardless of the PNR. However, with the increase of $T_{\mathrm{G}}$, the performance is rapidly improved, and then reaches the gentle descent region for all PNRs. 


Next, Fig. \ref{fig:simulation1}(b) demonstrates the average NMSE performance of all UEs in the small timescale, which is defined as $\mathrm{NMSE}_\mathrm{avg}=\mathbb{E}\Big\{\frac{1}{K}\sum_{k=1}^K\Big(\big\|\big(\mathbf{H}_k^T\otimes\mathbf{G}\big)-\big(\widehat{\mathbf{H}}_k^T\otimes\widehat{\mathbf{G}}\big)\big\|^2_{F}\big/\big\|\mathbf{H}_k^T\otimes\mathbf{G}\big\|^2_F \Big)\Big\}$, based on the $\widehat{\mathbf{G}}$ obtained in the large timescale with $T_{\mathrm{G}}=900$. The PNR is set the same in both the large and small timescales. 
As can be observed from this figure, in this step the NMSE decreases monotonously with the increase of $T_{\mathrm{H}}$ or PNR, which is similar to that in the large timescale estimation shown in Fig. \ref{fig:simulation1}(a). However, the required $T_{\mathrm{H}}$ is much smaller than $T_{\mathrm{G}}$ to achieve similar NMSE performance, e.g., $T_{\mathrm{H}}=120$ and $T_{\mathrm{G}}=500$ to reach a NMSE of $10\mathrm{dB}$ when $\mathrm{PNR}=20\mathrm{dB}$. This is mainly because in the small timescale, the BS only needs to estimate the UE-BIOS channel $\mathbf{H}_k$ for each UE with the obtained $\widehat{\mathbf{G}}$, while in the large timescale, both ${\mathbf{G}}$ and $\mathbf{H}$ should be estimated for the selected UE$_{\mathrm{fra}}$. However, even if the $T_{\mathrm{H}}$ is sufficient, the NMSE performance in the small timescale cannot exceed that in the large timescale estimation shown in Fig. \ref{fig:simulation1}(a), since the channel estimation performance of small timescale is limited by the estimation quality of $\widehat{\mathbf{G}}$ obtained in the large timescale. 

\subsection{Sum Rate Performance}
\begin{figure}
	\centering
	\includegraphics[height=7.9cm,width=14.9cm]{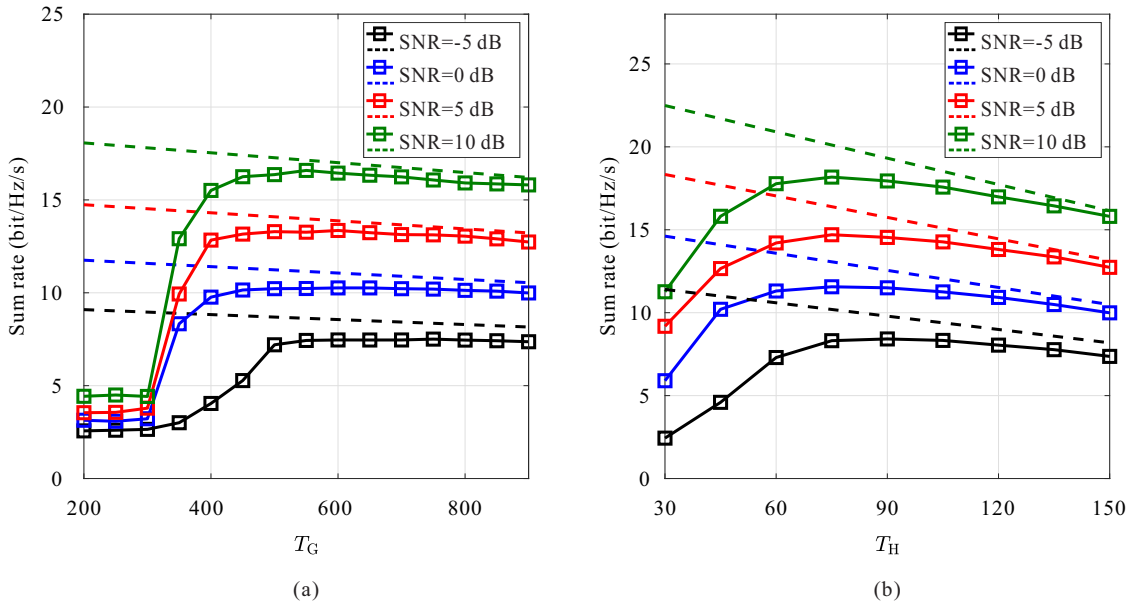}
	\caption{The sum rate performance of BIOS-assisted system with (solid lines) or without (dashed lines) estimation errors. (a) Sum rate versus $T_{\mathrm{G}}$; (b) sum rate versus $T_{\mathrm{H}}$.}	\label{fig:simulation1.5}
\end{figure}
To further evaluate the effectiveness of the proposed HTT-MO channel estimation scheme and the WMMSE-CD beamforming scheme, Fig. \ref{fig:simulation1.5} exhibits the sum rate performance of the BIOS-assisted system with or without channel estimation errors, represented by the solid and dashed lines, respectively. In the case with channel estimation errors, the CSI used for the beamforming is obtained by the HTT-MO scheme with $\mathrm{PNR}=\mathrm{SNR}+10 \mathrm{dB}$. In the case without channel estimation errors, the pilot overhead is still counted and set the same as that in the case with estimation errors in order to provide a benchmark. 


Fig. \ref{fig:simulation1.5}(a) shows the sum rate performance as a function of $T_\mathrm{G}$, when $T_\mathrm{H}$ is fixed at $150$. It can be seen that the sum rate performance is very poor due to the low channel estimation quality when $T_\mathrm{G}$ is very insufficient (below $300$), and starts to increase rapidly when $T_\mathrm{G}$ becomes larger. As $T_{\mathrm{G}}$ continues to increase, however, the sum rate begins to decrease as the estimation quality cannot be significantly improved, while according to (\ref{eqn:efficient SE}), the sum rate gradually decreases with more pilot overhead. In general, there is a best choice of $T_{\mathrm{G}}$ to balance the sum rate improvement due to more accurate channel estimation quality and the data rate loss due to more pilot overhead. 

Fig. \ref{fig:simulation1.5}(b) also provides the sum rate performance as a function of $T_\mathrm{H}$ with $T_\mathrm{G}=900$. Similar to that in Fig. \ref{fig:simulation1.5}(a), the sum rate with estimated CSI first increases and then decreases with the increase of $T_\mathrm{H}$, reaching the maximum sum rate with about $T_\mathrm{H}=75$ due to the trade-off between the estimation accuracy and the consumption of training overhead.

\subsection{Comparison of the Sum Rate Performance of BIOS-, IRS- and IOS-assisted systems}
\begin{figure}
	\centering
	\includegraphics[height=8cm,width=14.9cm]{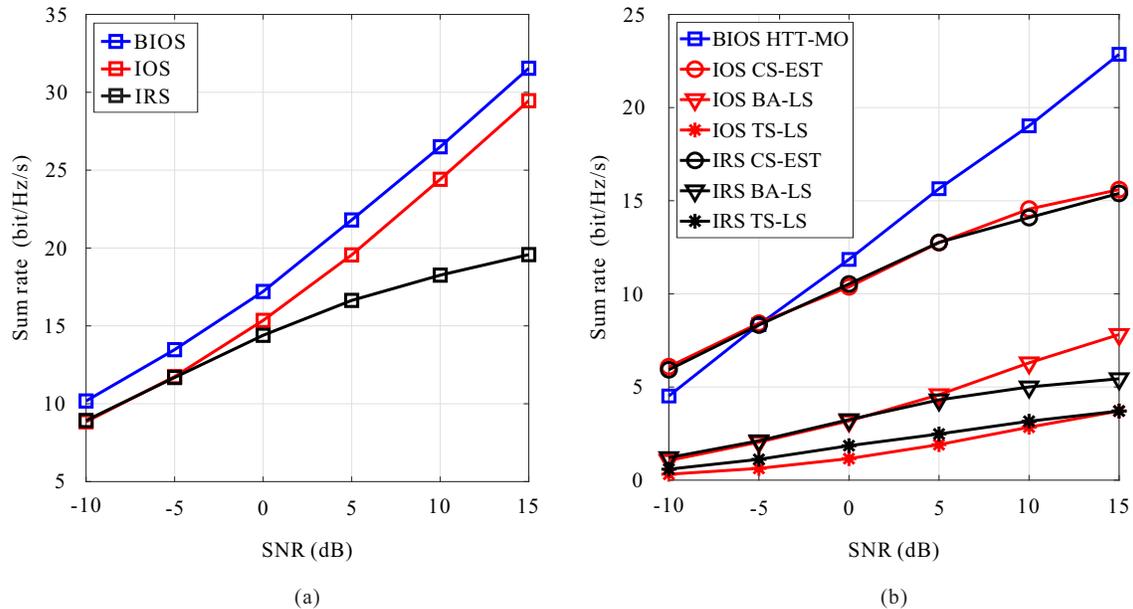}
	\caption{Sum rate versus SNR for BIOS-, IRS- and IOS-assisted systems. (a) With perfect CSI; (b) with estimated CSI.}	\label{fig:simulation3}
\end{figure}
In this subsection, we compare the sum rate performance of the proposed BIOS-assisted system to that of the conventional IRS- and IOS-assisted systems with both perfect CSI and estimated CSI. For the IOS-assisted system, we set $\epsilon=0.5$ to simultaneously serve UEs on both sides. The beamforming design of all these three systems are accomplished by the proposed WMMSE-CD scheme as both IRS and IOS can be regarded as a special case of BIOS.

Fig. \ref{fig:simulation3}(a) first depicts the sum rate performance of the BIOS-, IRS- and IOS-assisted systems versus SNR with perfect CSI, where the BS is assumed to have perfect knowledge of CSI without any pilot consumption. As can be observed, for all of the three systems, the sum rate performance increases monotonously with the SNR, while the BIOS always outperforms the IRS and IOS as it can independently control the beamforming on both sides. Meanwhile, the IRS performs the worst as it can only serve the UEs on the same side with the BS. 

Next, we compare their performance with estimated CSI. For the channel estimation in the conventional IRS- and IOS-assisted systems, we adopt the following channel estimation schemes.
\begin{itemize}
\item TS-LS \cite{STAR-RIS_channel_estimation_2021}: The authors of \cite{STAR-RIS_channel_estimation_2021} directly estimated the cascaded channels of all UEs one-by-one over consecutive time slots based on the LS algorithm for the IOS-assisted SISO system. We extend this scheme to the IOS- and IRS-assisted MIMO systems in our simulation.
\item BA-LS \cite{BALS}: Instead of directly estimating the cascaded channels, the authors of \cite{BALS} decoupled the IRS-BS and UE-IRS channels by modeling the received pilots as a tensor, and then estimated these channels with an alternating LS algorithm for the IRS-assisted MIMO system. We extend this scheme to the IOS-assisted MIMO system in our simulation.
\item CS-EST \cite{sparse_property_channel_estimation_2022}:  A three-stage scheme was proposed in \cite{sparse_property_channel_estimation_2022} to estimate the cascaded channel for the IRS-assisted single-user MIMO system. The AoDs at the UE, AoAs at the BS and the cascaded channel are estimated in each stage by the OMP algorithm. In this paper, we apply this scheme in the IOS- and IRS-assisted multi-user MIMO systems by separately estimating the cascaded channels of all UEs.
\end{itemize}

Fig. \ref{fig:simulation3}(b) depicts the sum rate performance of the BIOS-, IRS- and IOS-assisted systems as a function of SNR with estimated CSI. The PNR is assumed to be $10\mathrm{dB}$ higher than the SNR. As more training overhead provides higher estimation accuracy but occupies more transmission time, we select the best overhead for each estimation scheme and for each SNR in the sense of maximizing the sum rate of the corresponding system, instead of setting a fixed number of training pilots. 

By comparing Fig. \ref{fig:simulation3}(b) with Fig. \ref{fig:simulation3}(a), we can see that all of the three systems suffer from rate reduction due to the consumption of training overhead as well as channel estimation errors. As the TS-LS scheme does not exploit any property of channels in the IRS- and IOS-assisted systems, it needs extremely high training overhead, e.g., at least $KN_{\mathrm{UE}}M$ pilot vectors in a small timescale, and thus reaches the smallest sum rate for both IRS- and IOS-assisted systems. For the BA-LS scheme, although it utilizes the property that all UEs share the same RIS-BS channel, the required number of training overhead is still high, e.g., at least $KN_{\mathrm{UE}}(M+1-P)$ pilot vectors in a small timescale, which significantly affects the sum rate performance of IRS- and IOS-assisted systems. By contrast, the CS-EST scheme can efficiently exploit the angle domain sparsity of channels in the IRS- and IOS-assisted systems, the required training overhead in a small timescale is only in the order of $\mathcal{O}\left(KPQ\mathrm{log}(PQG_\mathrm{I})\right)$. Thus, the sum rate of both the IRS- and IOS-assisted systems with the CSI obtained by the CS-EST scheme is much higher than that with the TS-LS and BA-LS schemes. Finally, although the channel estimation in the BIOS-assisted system is much more complicated than that in the IRS- and IOS-assisted systems as we discussed in Section \ref{sec:system-model-and-channel-model} and Section \ref{sec:heterogeneous-channel-estimation-strategy}, results in this figure show that the BIOS-assisted system with the CSI obtained by the proposed HTT-MO scheme significantly outperforms the IRS- and IOS-assisted systems for medium and high SNRs. This performance advantage not only results from the ability of flexible beamforming control on both sides bestowed by the BIOS structure, but also results from the reduction of requested training overhead by efficiently exploiting the channel properties in the BIOS-assisted system, i.e., the two-timescale, heterogeneous and sparsity properties.

\section{Conclusion}\label{sec:conclusion}
In this paper, we have investigated the channel estimation problem of the uplink BIOS-assisted multi-user MIMO system. To reduce the large pilot overhead in the channel estimation of the BIOS-assisted system, we proposed the HTT-MO channel estimation scheme by efficiently exploiting the heterogeneous, two-timescale, and sparsity channel properties, in which the BS first estimates the common BIOS-BS channel with the pilots sent by a selected UE$_{\mathrm{fra}}$ in a large timescale, and then uses the estimated BIOS-BS channel to estimate the UE-BIOS channels for all UEs separately in every small timescale. In each step, by exploiting the low rank property due to the channel sparsity and applying the MO method to deal with the rank constraint, the requested pilot overhead is further reduced. In addition, we have proposed the WMMSE-CD scheme for the beamforming optimization of the downlink BIOS-assisted system to maximize the UEs' sum rate. We have provided various simulation results to demonstrate the effectiveness of the proposed channel estimation and beamforming schemes. It has been shown that compared with the conventional RISs such as IRS and IOS, the BIOS-assisted system with the CSI estimated by the proposed HTT-MO scheme has remarkable performance advantage in sum rate, not only resulted from the truth that the BIOS can provide very flexible beamforming on both sides, but also caused by the efficient utilization of the channel properties in the HTT-MO channel estimation scheme. 

\begin{appendices}
\section{Proofs of \eqref{eqn:set of G and H fle2} and \eqref{eqn:set of G and H fra2} in Lemma \ref{lem:freedom of G and H}}\label{app:equivalence between sets}
The equivalence between the two forms of $\mathcal{S}_{\mathrm{fra}, k}$ in (\ref{eqn:set of G and H for fra}) and (\ref{eqn:set of G and H fra2}) can be proved by showing that for any matrices $\mathbf{A},\mathbf{C}\in\mathbb{C}^{M_1\times N_1}$, $\mathbf{B},\mathbf{D}\in\mathbb{C}^{M_2\times N_2}$ without any zero element, the necessary and sufficient condition of $\mathbf{A}\otimes\mathbf{B}=\mathbf{C}\otimes\mathbf{D}$ is: $\exists a \neq 0, \mathbf{A} = a\mathbf{C}, \mathbf{B} = \frac{1}{a}\mathbf{D}$. 

\textbf{Necessity}: If there is an $a\neq0$, $\mathbf{A} = a\mathbf{C}, \mathbf{B} = \frac{1}{a}\mathbf{D}$, we have $\mathbf{A}\otimes\mathbf{B}=(a\times\frac{1}{a})\mathbf{C}\otimes\mathbf{D}=\mathbf{C}\otimes\mathbf{D}$ with the property of Kronecker product: $\forall a,b\neq0, \ a\mathbf{A}\otimes b\mathbf{B} = ab (\mathbf{A}\otimes\mathbf{B})$.

\textbf{Sufficiency}: $\forall m_1, n_1, m_2, n_2$, we have:
\begin{equation}
     [\mathbf{A}]_{m_1n_1}[\mathbf{B}]_{m_2n_2} =     [\mathbf{A}\otimes\mathbf{B}]_{((m_1-1)M_2+m_2)((n_1-1)N_2+n_2)}\neq0,
\end{equation}
\begin{equation}
     [\mathbf{C}]_{m_1n_1}[\mathbf{D}]_{m_2n_2} = [\mathbf{C}\otimes\mathbf{D}]_{((m_1-1)M_2+m_2)((n_1-1)N_2+n_2}\neq0.
\end{equation}
As $\mathbf{A}\otimes\mathbf{B}=\mathbf{C}\otimes\mathbf{D}$, each element of $\mathbf{A}\otimes\mathbf{B}$ should be equal to the corresponding element of $\mathbf{C}\otimes\mathbf{D}$, i.e., $[\mathbf{A}]_{m_1n_1}[\mathbf{B}]_{m_2n_2} = [\mathbf{C}]_{m_1n_1}[\mathbf{D}]_{m_2n_2}$ for any $m_1$, $n_1$, $m_2$ and $n_2$. By setting $a=\frac{[\mathbf{A}]_{11}}{[\mathbf{C}]_{11}}$, $[\mathbf{C}]_{11}$ and $[\mathbf{D}]_{11}$ can thus be expressed as $[\mathbf{C}]_{11}=\frac{1}{a}[\mathbf{A}]_{11}$, $[\mathbf{D}]_{11}=a[\mathbf{B}]_{11}$. Therefore, for all $m$ and $n$, $[\mathbf{A}]_{11}[\mathbf{B}]_{mn} = [\mathbf{C}]_{11}[\mathbf{D}]_{mn} = \frac{1}{a}[\mathbf{A}]_{11}[\mathbf{D}]_{mn}$, which means that $[\mathbf{B}]_{mn} =\frac{1}{a}[\mathbf{D}]_{mn}$ holds for all $m$ and $n$. Thus, we have $\mathbf{B} = \frac{1}{a}\mathbf{D}$. Similarly, we can prove that $\mathbf{A} = a\mathbf{C}$, which completes the proof of sufficiency.$\hfill\blacksquare$

To show the equivalence between the two forms of $\mathcal{S}_{\mathrm{fle}, k}$ in (\ref{eqn:set of G and H for fle}) and (\ref{eqn:set of G and H fle2}), as the Khatri-Rao product can be divided into the Kronecker product of each column of the two matrices, the proof is thus similar to that above. 
\end{appendices}

\end{document}